\documentclass[transmag,journal]{IEEEtran}
\usepackage[T1]{fontenc}
\usepackage[latin9]{inputenc}
\usepackage{color}
\usepackage{textcomp}
\usepackage{amsbsy}
\usepackage{amstext}
\usepackage{amsmath}
\usepackage{graphicx}
\usepackage[unicode=true,
 bookmarks=true,bookmarksnumbered=true,bookmarksopen=true,bookmarksopenlevel=1,
 breaklinks=false,pdfborder={0 0 0},pdfborderstyle={},backref=false,colorlinks=false]
 {hyperref}
\hypersetup{pdftitle={Your Title},
 pdfauthor={Your Name},
 pdfpagelayout=OneColumn, pdfnewwindow=true, pdfstartview=XYZ, plainpages=false}

\usepackage{fancyhdr}

\makeatletter

\newcommand{\lyxmathsym}[1]{\ifmmode\begingroup\def\b@ld{bold}
  \text{\ifx\math@version\b@ld\bfseries\fi#1}\endgroup\else#1\fi}

\let\oldforeign@language\foreign@language
\DeclareRobustCommand{\foreign@language}[1]{%
  \lowercase{\oldforeign@language{#1}}}

\usepackage[caption=false,font=normalsize,labelfont=sf,textfont=sf]{subfig}

\makeatother

\begin{document}
\title{\textcolor{black}{Exceptional Degeneracies in Traveling Wave Tubes
with Dispersive Slow-Wave Structure Including Space-Charge Effect}}
\author{\IEEEauthorblockN{Kasra Rouhi\IEEEauthorrefmark{1}, Robert Marosi\IEEEauthorrefmark{1},
Tarek Mealy\IEEEauthorrefmark{1}, Ahmed F. Abdelshafy\IEEEauthorrefmark{1},
Alexander Figotin\IEEEauthorrefmark{2}, and Filippo Capolino\IEEEauthorrefmark{1}}\IEEEauthorblockA{\IEEEauthorrefmark{1}Department of Electrical Engineering and Computer
Science, University of California, Irvine, CA 92697 USA\\
}\IEEEauthorblockA{\IEEEauthorrefmark{2}Department of Mathematics, University of California,
Irvine, CA 92697 USA\\
}}

\IEEEtitleabstractindextext{
\begin{abstract}
The interaction between a linear electron beam and a guided electromagnetic
wave is studied in the contest of exceptional points of degeneracy
(EPD) supported by such an interactive system. The study focuses on
the case of a linear beam traveling wave tube (TWT) with a realistic
helix waveguide slow-wave structure (SWS). The interaction is formulated
by an analytical model that is a generalization of the Pierce model,
assuming a one-dimensional electron flow along a dispersive single-mode
guiding SWS and taking into account space-charge effects in the system.
The augmented model using phase velocity and characteristic impedance
obtained via full-wave simulations is validated by calculating gain
versus frequency and comparing it with that from more complex electron
beam simulators. This comparison also shows the accuracy of our new
model compared with respect to the non-dispersive Pierce model. EPDs
are then investigated using the augmented model, observing the coalescence
of complex-valued wavenumbers and the system's eigenvectors. The point
in the complex dispersion diagram at which the TWT-system starts/ceases
to exhibit a convection instability, i.e., a mode starts/ceases to
grow exponentially along the TWT, is the EPD. We also demonstrate
the EPD existence by showing that the Puiseux fractional power series
expansion well approximates the bifurcation of the dispersion diagram
at the EPD. This latter concept also explains the \textquotedbl exceptional\textquotedbl{}
sensitivity of the TWT-system to changes in the beam's electron velocity
when operating near an EPD.
\end{abstract}

}
\maketitle
\IEEEPARstart{H}{igh}\  power traveling wave tube (TWT) amplifiers
are of high importance for telecommunications, high-performance radar
applications, including atmospheric studies, precision tracking, and
high-resolution imaging \cite{shiffler1989high,hung2012high,armstrong2018compact}.
A TWT uses a slow-wave structure (SWS) as a key component to harvest
energy from an electron beam (e-beam) into radio frequency waves efficiently
over broad bandwidths \cite{naqvi1996axial,qiu2009vacuum}. In this
paper, we focus on a realistic helix TWT with dispersive SWS's characteristic
parameters. The interaction with the e-beam affects the way EM waves
propagate in the so-called \textquotedbl hot\textquotedbl{} circuit,
i.e., accounting for the beam-electromagnetic (EM) mode interaction.

An exceptional point of degeneracy (EPD) in a system refers to the
property of the relevant system matrix that contains at least one
nontrivial Jordan block structure, i.e., when two or more eigenvectors
coalesce into a single degenerate one \cite{heiss2000repulsion,Heiss2004Exceptional,Figotin2005Gigantic,Othman2017Theory,Rouhi2020Exceptional}.
The concept of EPD has been studied in lossless, spatially \cite{Figotin2003Oblique,Othman2016GiantGain},
or temporally \cite{Kazemi2019Exceptional,kazemi2020ultra} periodic
structures, and in systems with loss and/or gain under parity-time
symmetry \cite{Bender1998Real,Heiss2004Exceptional,Schindler2011Experimental}.
We employ the Puiseux fractional power expansion series to illustrate
the bifurcation of the system's dispersion diagram at the EPD \cite{Kato1995Perturbation}.
The EPD has been studied for its applications in sensing devices \cite{Wiersig2016Sensors,Chen2018Generalized}
and oscillators \cite{Hodaei2014Parity,Othman2016Low,Veysi2018Degenerate,abdelshafy2020distributed}.

\begin{figure}[t]
\begin{centering}
\includegraphics[width=3.5in]{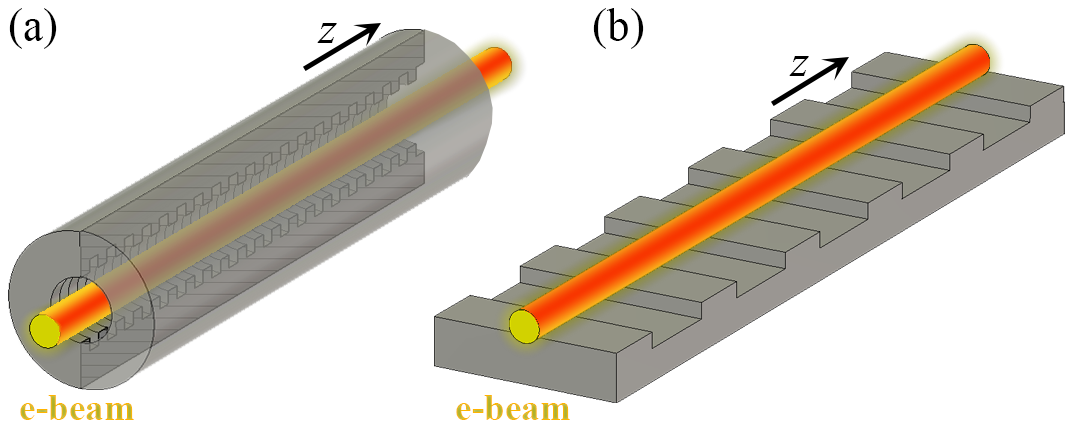}
\par\end{centering}
\centering{}\caption{Illustrative schematics showing an e-beam in the proximity of an EM-guiding
SWS. (a) Beam inside a circular waveguide with corrugations, and (b)
beam near a periodic grating.\label{fig: Schematics}}
\end{figure}

\begin{figure*}[t]
\begin{centering}
\includegraphics[width=7in]{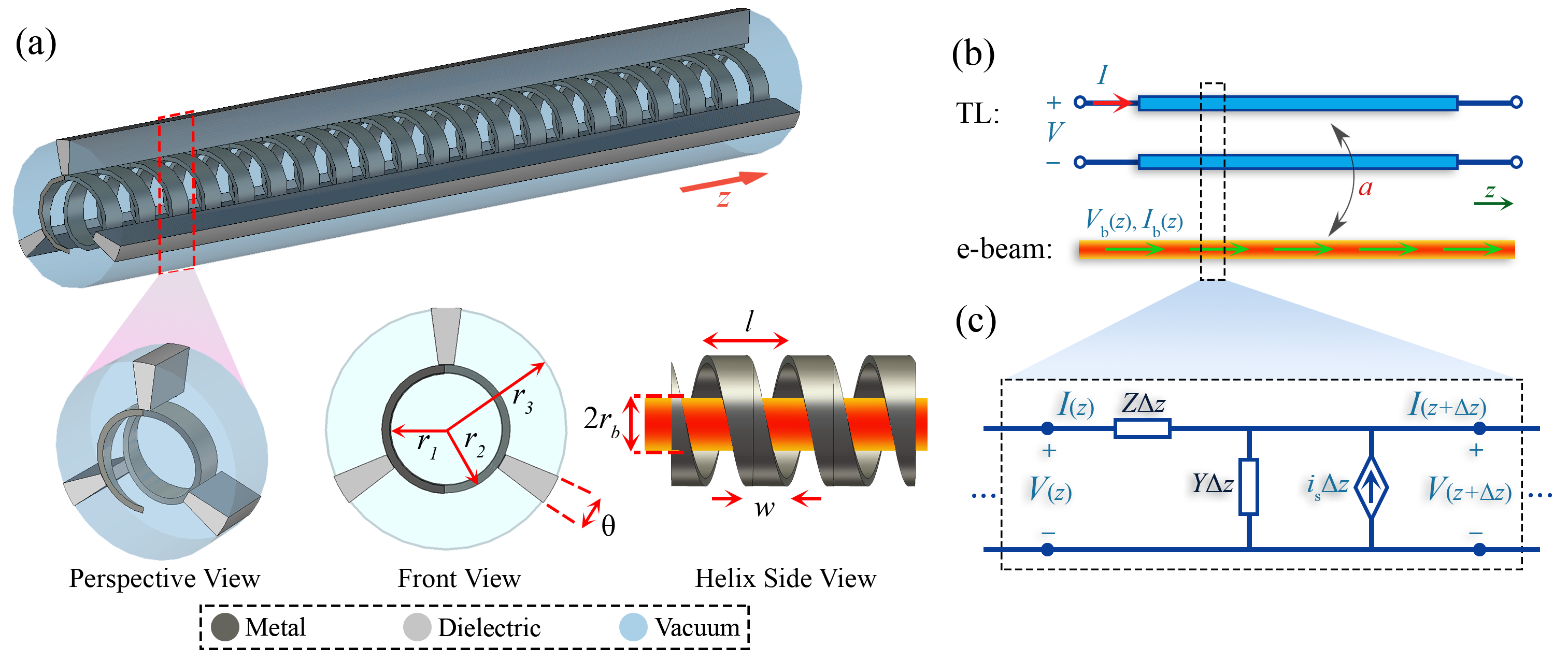}
\par\end{centering}
\centering{}\caption{(a) Tape helix SWS in a circular metallic waveguide with radius $r_{3}=1.06\:\mathrm{mm}$.
An e-beam with radius $r_{b}=560\:\mathrm{\mu m}$ flows along the
axis of the helical conductor of inner radius $r_{1}=744\:\mathrm{\mu m}$,
and outer radius $r_{2}=846\:\mathrm{\mu m}$ supported by dielectric
rods. The other geometric parameters are $l=1.04\:\mathrm{mm}$, $w=520\:\mathrm{\mu m}$,
and $\theta=14.2\lyxmathsym{\protect\textdegree}$. (b) Schematic
of the equivalent TL coupled to the e-beam used to study the hybrid
EM-space charge wave. (c) Equivalent TL circuit showing the per-unit-length
impedance, admittance and current generator $i_{s}$ that represents
the effect of the e-beam on the TL. \label{fig: Main}}
\end{figure*}

\begin{figure*}[t]
\begin{centering}
\includegraphics[width=5.8in]{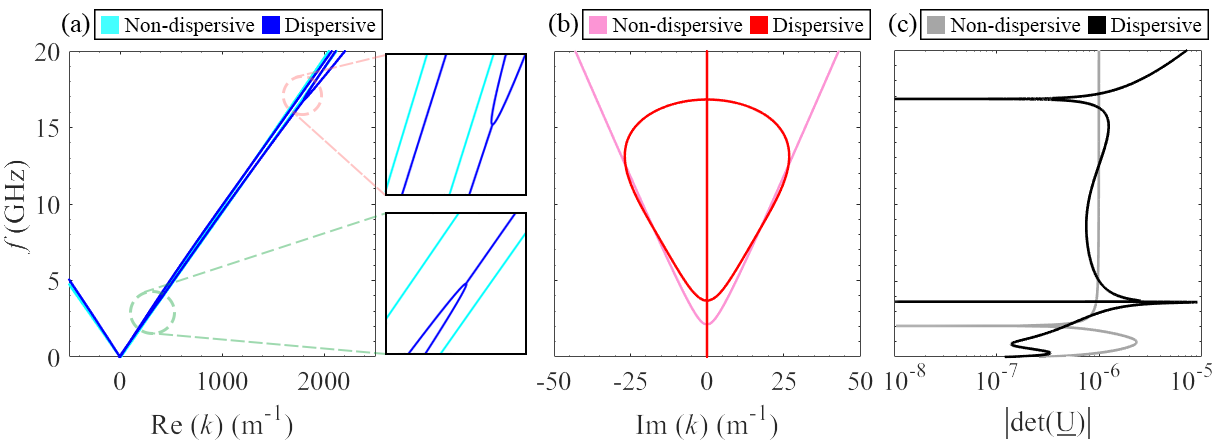}
\par\end{centering}
\centering{}\caption{Dispersion diagram of the four complex-valued wavenumbers versus frequency,
which shows two (one) bifurcation points correspond to the EPDs in
the dispersive (non-dispersive) system, and $\left|\det(\mathbf{\underline{U}})\right|$
whose minima indicate the occurrence of EPDs.\label{fig: DispersionOmega}}
\end{figure*}

In \cite{mealy2020exceptional}, the EPD in a system of an e-beam
interacting with an EM mode guided in a non-dispersive SWS with distributed
power extraction and without accounting for electrons' debunching
is used to conceive an effective oscillator. The mathematical formulation
in \cite{figotin2020exceptional} does not include waveguide dispersion
in the model, and above all, it cannot capture the second EPD that
occurs in the real TWTs at a higher frequency (See Fig. \ref{fig: DispersionOmega}).
This paper explains the fundamental physics describing the TWT operation,
accounting for electrons' debunching caused by space-charge effects
and SWS frequency dispersion, and describes the bifurcation points
in the dispersion diagram using the coalescence of the system's eigenvectors.
Also, the EPD-related fractional power expansion is used to explain
the wavenumber's extreme sensitivity to perturbations to the system
parameters such as operating frequency and e-beam's velocity and how
EPDs are related to the TWT amplification bandwidth. We calculate
the characteristic parameters of the cold guiding SWS based on what
Pierce proposed in \cite{pierce1947theory,Pierce1951waves,Pierce1954Traveling}
and further developed for frequency-dependent TWT-systems in analogy
(but differently) to what done in \cite{wohlbier2002multifrequency,converse2004impulse}.

The system consists of an EM field in a guiding SWS interacting with
a single e-beam flowing in the $z$-direction is schematically shown
in Fig. \ref{fig: Schematics}. The e-beam's electrons have average
velocity and linear charge density $u_{0}$ and $\rho_{0}$, respectively.
The e-beam has an average current $I_{0}=-\rho_{0}u_{0}$ and an equivalent
kinetic non-relativistic d.c. voltage $V_{0}=u_{0}^{2}/2\eta$, where
$\eta=e/m=1.758820\times10^{11}\:\mathrm{C/Kg}$ is the charge-to-mass
ratio of the electron with charge $-e$ and rest mass $m$. The small-signal
modulation in the e-beam velocity and charge density $u_{b}$ and
$\rho_{b}$, respectively, describe the so-called \textquotedblleft space-charge
wave\textquotedblright . The a.c. beam current and equivalent voltage
are given by $i_{b}=u_{b}\rho_{0}+u_{0}\rho_{b}$ and $v_{b}=u_{b}u_{0}/\eta$,
where we have retained only the linear terms based on the small-signal
approximation \cite{Pierce1951waves}, as also explained in Appendix A. We implicitly assume a time dependence of $\exp(j\omega t)$,
so the a.c. space-charge wave modulating the e-beam is described in
the phasor domain with $V_{b}(z)$ and $I_{b}(z)$, as

\begin{equation}
\partial_{z}V_{b}=-j\beta_{0}V_{b}-aZI-j\frac{I_{b}}{A\varepsilon_{0}\omega},\label{eq: V_bm}
\end{equation}

\begin{equation}
\partial_{z}I_{b}=-jgV_{b}-j\beta_{0}I_{b},\label{eq: I_bm}
\end{equation}
where $\beta_{0}=\omega/u_{0}$ is the phase constant of the space-charge
wave (when neglecting plasma frequency effects), $g=I_{0}\beta_{0}/(2V_{0})$,
$Z$ is the equivalent TL distributed series impedance, and $I(z)$
is the equivalent TL current, as will be explained later. Furthermore,
$E_{z}=E_{w}+E_{p}=aZI+jI_{b}/(A\varepsilon_{0}\omega)$ is the longitudinal
polarization (in the $z$-direction) of the electric field component
that modulates the velocity and bunching of the electrons. The longitudinal
field $E_{z}=E_{w}+E_{p}$ is the sum of two components. The term
$E_{p}$ accounts for nonuniform charge density, and in the phasor
form is given by $E_{p}=jI_{b}/(A\varepsilon_{0}\omega)$ \cite[Chapter 10]{gewartowski1965principles},
where $A$ is the e-beam cross-sectional area, and $\varepsilon_{0}$
is vacuum permittivity. The term $E_{p}$ is generated by charge distribution
that also causes the so-called ``debunching'', and its calculation
is in agreement with the Lagrangian model for TWT-systems in \cite[Chapter 7]{figotin2020analytic},
as explained in Appendix G. In addition, the term $E_{w}=aZI$
is the longitudinal electric field of the EM mode propagation in the
SWS, affecting the bunching of the e-beam, according to the well-known
Pierce model \cite{Pierce1951waves}. Also, the term $a$ represents
a coupling strength coefficient, that describes how the e-beam couples
to the TL as already introduced in \cite{tamma2014extension} and
\cite[Chapter 3]{figotin2013multi,figotin2020analytic}. More details
on the fundamental equations describing the interacting system are
in Appendix B. The well-known telegrapher\textquoteright s
equations describe the EM modal propagation in the SWS, based on the
equivalent TL model shown in Figs. \ref{fig: Main}(b) and (c) where
the distributed per-unit-length series impedance $Z$, and shunt admittance
$Y$ relate the equivalent TL voltage $V(z)$ and current $I(z)$
phasors as

\begin{equation}
\partial_{z}V=-ZI,\label{eq: TelegVm}
\end{equation}

\begin{equation}
\partial_{z}I=-YV-a\partial_{z}I_{b}.\label{eq: TelegIm}
\end{equation}

Here, the term $i_{s}=-a\partial_{z}I_{b}$ represents a distributed
current generator \cite{marcuvitz1951representation}, that accounts
for the effect of the beam's charge wave flowing in the SWS \cite{Pierce1951waves,tamma2014extension}.
In order to construct an accurate model that provides precise predictions
for realistic structures and overcome the simplicity of the ideal
assumptions in the original Pierce model, we use frequency-dependent
waveguide parameters in the equations. In practice, we first analyze
wave propagation in the ``cold'' SWS (i.e., in the absence of the
e-beam) using a full-wave method to get the values of $Z(\omega)$
and $Y(\omega)$ to be used in the formulation. To recover these frequency-dependent
characteristic parameters, we use the finite element method eigenmode
solver in CST Studio Suite and extract the cold circuit EM phase velocity
$v_{c}(\omega)=\omega/\beta_{c}(\omega)$, where $\beta_{c}(\omega)=\sqrt{-Z(\omega)Y(\omega)}$
is the phase propagation constant of the cold SWS mode, and the equivalent
TL characteristic impedance $Z_{c}(\omega)$. By using the extracted
values for $v_{c}(\omega)$ and $Z_{c}(\omega)$, the equivalent frequency-dependent
distributed series impedance $Z(\omega)=j\omega Z_{c}(\omega)/v_{c}(\omega)=jZ_{c}(\omega)\beta_{c}(\omega)$
and shunt admittance $Y(\omega)=j\omega/\left(Z_{c}(\omega)v_{c}(\omega)\right)=j\beta_{c}(\omega)/Z_{c}(\omega)$
are calculated. Moreover, in Eq. (\ref{eq: TelegIm}), when $a=0$
the e-beam is not coupled to the TL, and when $a=1$, the model reduces
to the one developed in \cite{Pierce1951waves,Pierce1954Traveling}.
The presence of this coupling strength coefficient generalizes what
was done in \cite{Pierce1951waves,Pierce1954Traveling}, since the
beam may be subject to a strong longitudinal electric field that is
not accurately accounted for by the simple circuit impedance $Z_{c}$
of the originally Pierce model \cite{Pierce1951waves,Pierce1954Traveling}.
For convenience, we define a state vector $\boldsymbol{\Psi}(z)=[V,I,V_{b},I_{b}]^{T}$
that describes the hybrid EM-space charge wave propagation along $z$,
and rewrite Eqs. (\ref{eq: V_bm}), (\ref{eq: I_bm}), (\ref{eq: TelegVm}),
and (\ref{eq: TelegIm}) in matrix form as

\begin{equation}
\partial_{z}\boldsymbol{\Psi}(z)=-j\mathbf{\underline{M}}\boldsymbol{\Psi}(z),\label{eq: System Equationm}
\end{equation}
where $\underline{\mathbf{M}}$ is the $4\times4$ system matrix that
after replacing $v_{c}(\omega)$ and $\beta_{c}(\omega)$ in the system
equations, reads as

\begin{equation}
\underline{\mathbf{M}}=\left[\begin{array}{cccc}
0 & \beta_{c}(\omega)Z_{c}(\omega) & 0 & 0\\
\beta_{c}(\omega)/(Z_{c}(\omega)) & 0 & -ag & -a\beta_{0}\\
0 & a\beta_{c}(\omega)Z_{c}(\omega) & \beta_{0} & R_{p}\\
0 & 0 & g & \beta_{0}
\end{array}\right].
\end{equation}

In the above system matrix, $R_{p}$ is a space-charge parameter related
to the debunching of beam's charges, and is given by \cite{Othman2016Giant,Othman2016Theory}

\begin{equation}
R_{p}=\frac{1}{A\varepsilon_{0}\omega}=\frac{2V_{0}\omega{}_{q}^{2}}{\omega I_{0}u_{0}},
\end{equation}
where $\omega_{q}=R_{sc}\omega_{p}$ is the reduced plasma angular
frequency, $\omega_{p}=\sqrt{-\rho_{0}\eta/(A\varepsilon_{0})}=\sqrt{I_{0}u_{0}/\left(2V_{0}A\varepsilon_{0}\right)}$
is the plasma frequency \cite{hammer1967coupling}, and $R_{sc}$
is the plasma frequency reduction factor \cite{booske2004insights}.
The term $R_{sc}$ accounts for reductions in the magnitude of the
axial component of the space-charge electric field due to either finite
beam radius or proximity of the surrounding conducting walls. Fields
produced by space-charges represent repulsive forces in a dense beam
of charged particles. Assuming a state vector $z$-dependence of the
form $\boldsymbol{\Psi}(z)\propto\exp(-jkz)$, where $k$ is the wavenumber
of a hybrid mode in the EM-space charge wave interacting system, the
eigenmodes are obtained by solving the eigenvalue problem $k\boldsymbol{\Psi}(z)=\mathbf{\underline{M}}\boldsymbol{\Psi}(z)$.
The resulting modal dispersion characteristic equation is given by

\begin{figure}[t]
\begin{centering}
\includegraphics[width=3in]{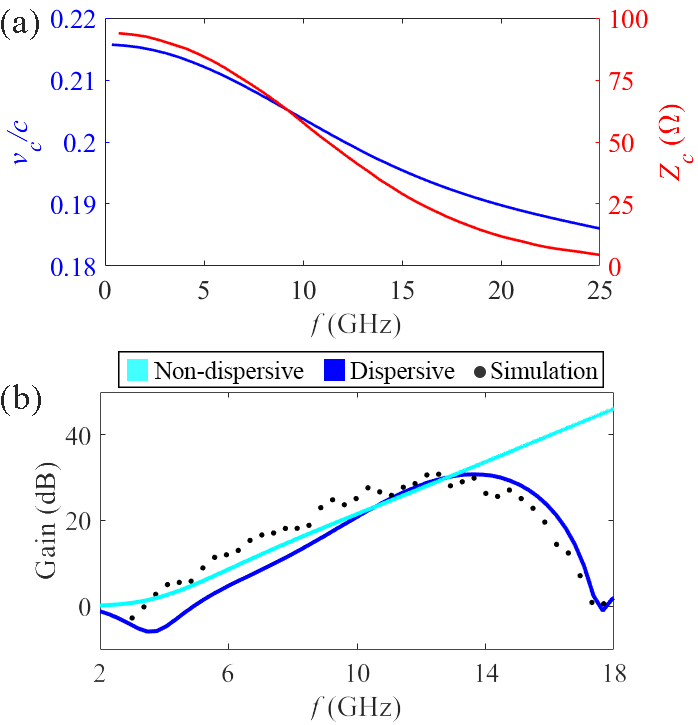}
\par\end{centering}
\centering{}\caption{(a) Phase velocity and Pierce impedance of the EM mode in the cold
SWS, obtained via full-wave eigenmode simulations. (b) Gain versus
frequency from the theoretical model based on the non-dispersive (cyan
curve), and dispersive (blue curve) solution of system in Eq. (\ref{eq: System Equationm}),
and simulation results using the software LATTE (black dots).\label{fig: HelixvcZcGain}}
\end{figure}

\begin{equation}
\begin{array}{c}
D(\omega,k)=\det(\mathbf{\underline{M}}-k\mathbf{\underline{I}})=k^{4}-k^{3}\left(2\beta_{0}\right)\\
\\
+k^{2}\left(\beta_{0}^{2}-\beta_{q}^{2}-\beta_{c}^{2}(\omega)+a^{2}g\beta_{c}(\omega)Z_{c}(\omega)\right)+k\left(2\beta_{c}^{2}(\omega)\beta_{0}\right)\\
\\
-\beta_{c}^{2}(\omega)\left(\beta_{0}^{2}-\beta_{q}^{2}\right)=0.
\end{array}\label{eq: Dispersionm}
\end{equation}
where $\beta_{q}=\omega_{q}/u_{0}=\sqrt{R_{p}g}$ is the phase constant
of space-charge wave traveling with a phase velocity equal to $u_{0}$
and at an angular frequency equal to $\omega_{q}$. The solution of
Eq. (\ref{eq: Dispersionm}) leads to four modal complex-valued wavenumbers
that describe the modes in the TWT interactive system. The above characteristic
equation is equivalently rewritten as

\begin{equation}
(k^{2}-\beta_{c}^{2})\left((k-\beta_{0})^{2}-\beta_{q}^{2}\right)=-a^{2}g\beta_{c}(\omega)Z_{c}(\omega)k^{2}.\label{eq: Dispersion2m}
\end{equation}

The first parenthesis on the left side of Eq. (\ref{eq: Dispersion2m})
only contains parameters related to TL and the second parenthesis
includes only the parameters of the e-beam. The term on the right
side represents the interaction between the EM mode in the SWS and
the e-beam, and it contains both e-beam and EM mode parameters. When
the e-beam and TL are decoupled, one has $a=0$, the two equations
in parenthesis become two independent dispersion equations.

A second-order EPD occurs when two eigenmodes coalesce in their eigenvalues
and eigenvectors. Thus, when such degeneracy occurs, the matrix $\mathbf{\mathbf{\underline{M}}}$
is similar to a matrix that contains a Jordan block of order two.
A necessary condition to have a second-order EPD is to have two repeated
eigenvalues, which means that the characteristic equation should have
two repeated roots as

\begin{equation}
D(\omega_{e},k)\propto(k-k_{e})^{2},\label{eq: Second-order EPDm}
\end{equation}
where $\omega_{e}$ and $k_{e}$ are the degenerate angular frequency
and wavenumber at the EPD. This happens when $D(\omega_{e},k_{e})=0$,
and$\left.\partial_{k}D(\omega_{e},k)\right|_{k=k_{e}}=0$. We derive
the following expressions for $Z=Z_{e}$ and $Y=Y_{e}$, which will
produce the EPD for given e-beam parameters

\begin{equation}
Z_{e}=\frac{j\left((k_{e}-\beta_{0})^{2}-R_{p}g\right)^{2}}{a^{2}g\left(-\beta_{0}^{2}+k_{e}\beta_{0}+R_{p}g\right)},\label{eq: Z_EPDm}
\end{equation}

\begin{equation}
Y_{e}=\frac{ja^{2}gk_{e}^{3}(k_{e}-\beta_{0})}{\left((k_{e}-\beta_{0})^{2}-R_{p}g\right)^{2}},\label{eq: Y_EPDm}
\end{equation}
where all the parameters are calculated at $\omega_{e}$ and $k_{e}$.
Assuming that the EPD conditions for impedance and admittance in Eqs.
(\ref{eq: Z_EPDm}) and (\ref{eq: Y_EPDm}) are satisfied, the degenerate
wavenumber $k_{e}$ is determined by the product of Eqs. (\ref{eq: Z_EPDm})
and (\ref{eq: Y_EPDm})

\begin{equation}
Z_{e}Y_{e}=\frac{-k_{e}^{3}(k_{e}-\beta_{0})}{(-\beta_{0}^{2}+k_{e}\beta_{0}+R_{p}g)}.
\end{equation}

We know that $\beta_{c,e}^{2}=-Z_{e}Y_{e}$ and $\beta_{q}^{2}=R_{p}g$
under the EPD condition, so we will calculate the wavenumber of the
degenerate hybrid mode $k_{e}$ by solving

\begin{equation}
\beta_{c,e}^{2}\beta_{q}^{2}=(k_{e}^{3}-\beta_{c,e}^{2}\beta_{0})(k_{e}-\beta_{0}).
\end{equation}

In order to investigate the EPD, we analytically derive the system's
eigenvector expressions related to the four wavenumbers $k_{n}$ with
$n=1,2,3,4$, that are written in the form of

\begin{equation}
\boldsymbol{\Psi}_{n}=\left[\begin{array}{c}
(k_{n}-\beta_{0})^{2}-\beta_{q}^{2}\\
\frac{k_{n}}{\beta_{c}Z_{c}}\left((k_{n}-\beta_{0})^{2}-\beta_{q}^{2}\right)\\
ak_{n}(k_{n}-\beta_{0})\\
agk_{n}
\end{array}\right].\label{eq: EigenVectorm}
\end{equation}

In summary, the two conditions in Eqs. (\ref{eq: Z_EPDm}) and (\ref{eq: Y_EPDm})
represent constraints on the TL parameters, calculated at the EPD
frequency. These two conditions need to be enforced to have a second-order
EPD, where two eigenmodes of the interacting system have identical
eigenvalues $k_{1}=k_{2}=k_{e}$ and eigenvectors $\boldsymbol{\Psi}_{1}=\boldsymbol{\Psi}_{2}=\boldsymbol{\Psi}_{e}$.
Suppose we find a set of parameters to satisfy the EPD condition;
in that case, these values lead to the same two eigenvalues and a
single corresponding eigenvector according to Eq. (\ref{eq: EigenVectorm}).

The helix SWS features a conventional two-body (input and output)
cylindrical vacuum envelope that contains a metallic tape helix supported
by three equally spaced dielectric rods, which are made of BeO with
$\varepsilon_{r}=6.5$ \cite{han2008thermal}. The SWS is illustrated
in Fig. \ref{fig: Main}(a), with the helix's geometric parameters
shown in the caption. Because the helix TWT dispersion is vital for
pulse amplification or nonstationary problems\textquoteright{} response
of the tube, the frequency dependence of the cold circuit phase velocity
and the interaction impedance must be included in the model \cite{wohlbier2002multifrequency,converse2004impulse,setayesh2017pawaic}.
We have simulated the helix SWS by using the finite element method
eigenmode solver in CST Studio Suite and extracted the characteristic
parameters, i.e., the cold circuit EM phase velocity $v_{c}(\omega)$,
and the equivalent TL characteristic impedance $Z_{c}(\omega)$; then
the calculated results is illustrated in Fig. \ref{fig: HelixvcZcGain}(a).
We demonstrate the occurrence of EPDs in the helix TWT using practical
values for both the helix SWS and e-beam. For the e-beam, we assume
$u_{0}=0.2c$, where $c$ is the speed of light, $I_{0}=47\:\mathrm{mA}$,
$V_{0}=10.5\:\mathrm{kV}$, and a e-beam radius equal to $r_{b}=560\:\mathrm{\mu m}$.
The resulting plasma frequency is $f_{p}=\omega_{p}/(2\pi)=624.6\:\mathrm{MHz}$.
We assume a plasma frequency reduction factor of $R_{sc}=0.12$, which
was calculated for the SWS in Fig. \ref{fig: Main}(a) using the software
LATTE \cite{antonsen1998traveling,wohlbier2002multifrequency,wohlbier2003latte}.
The maximum interaction between the space-charge wave and the EM wave
occurs when they are synchronized, i.e., by matching $v_{c}$ to $u_{0}$,
a condition that is specifically called \textquotedblleft synchronization\textquotedblright .
We calibrate the value of the coupling strength coefficient $a$,
which is an essential parameter of our model, to predict the gain
in the TWT; this coefficient can be used to calculate the gain of
longer TWT structures. In this example, we estimate $a=0.527$, as
extracted from simulations, and the required steps are explained in
Appendix E. Then, by solving the wavenumber dispersion
equation for specific frequency values, we obtain the real and imaginary
parts of the four hybrid wavenumbers in Fig. \ref{fig: DispersionOmega}(a)
and (b). Moreover, in Fig. \ref{fig: DispersionOmega}, we illustrate
the non-dispersive results by using an average value of $v_{c}$ and
$Z_{c}$ in the predetermined frequency range (See Fig. \ref{fig: HelixvcZcGain}(a)).
As we can observe, the non-dispersive model cannot capture the second
EPD at a higher frequency, which exists in realistic structures.

The similarity transformation matrix $\mathbf{\underline{U}}=[\mathrm{\mathbf{U}}_{1},\mathrm{\mathbf{U}}_{2},\mathrm{\mathbf{U}}_{3},\mathrm{\mathbf{U}}_{4}]$,
where the column $\mathbf{\underline{U}}_{i}$ is the eigenvector
corresponding to the $i$-th eigenvalue, diagonalizes the system matrix
as $\mathbf{\underline{M}}=\mathbf{\underline{U}\underline{\Lambda}\underline{U}^{\mathrm{-1}}}$.
At an EPD, at least two eigenvectors become linearly dependent, implying
that $\left|\mathrm{det}(\underline{\mathbf{U}})\right|$ vanishes.
In this example, we consider the EPD frequency of the dispersive model
at $f_{e}=16.82\:\mathrm{GHz}$ in the shown frequency range, where
$\beta_{0,e}=1762.21\:\mathrm{m}^{-1}$. In Fig. \ref{fig: DispersionOmega}(a)
and (b), we observe the bifurcation of the wavenumbers' real and imaginary
parts at the EPDs. The maximum TWT gain is at $f_{opt}=13.14\:\mathrm{GHz}$,
where the maximum value of the imaginary part occurs. This frequency
is close to the initial-design synchronization frequency $f_{sync}=12.10\:\mathrm{GHz}$,
where $v_{c}=u_{0}=0.2c$, as expected. It may be possible to shift
the maximum gain frequency by changing the frequencies of EPDs through
varying plasma frequency or other controllable parameters in the TWT-system.
In order to validate the proposed TWT model, which accounts for waveguide
mode dispersion and space-charge effect, we provide the gain versus
frequency plot obtained from simulations in LATTE and compare it to
the theoretically calculated gain based on applying boundary conditions
for the charge wave and EM mode as explained in Appendix E. We observe a good agreement between simulated and theoretical
results for the dispersive model in Fig. \ref{fig: HelixvcZcGain}(b),
which demonstrates the accuracy of the proposed method. This figure
shows that the non-dispersive model cannot predict the gain correctly
in the illustrated frequency range.

\begin{figure}[t]
\begin{centering}
\includegraphics[width=3.5in]{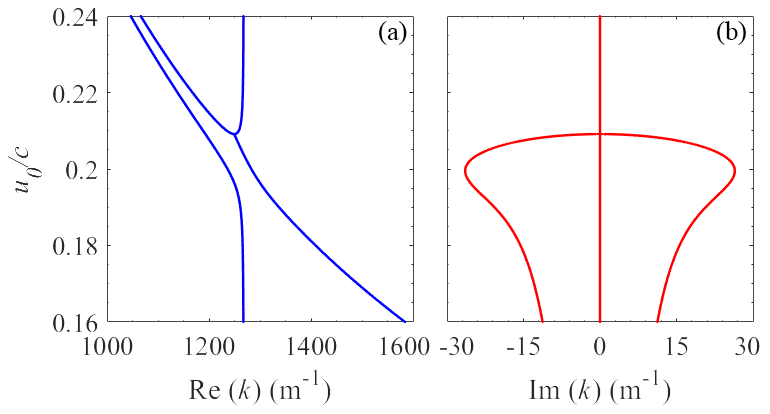}
\par\end{centering}
\centering{}\caption{Dispersion diagram of the three wavenumbers versus $u_{0}$. The diagram
shows a bifurcation point that corresponds to the EPD point, whereas
frequency is equal to the synchronization frequency $f_{sync}$.\label{fig: Dispersionu_0}}
\end{figure}

So far, we have analyzed the dispersion diagram by varying frequency.
In the next step, we investigate the wavenumber dispersion diagram
varying the electron's average velocity out of synchronization and
observe EPDs under these conditions. Hence, we assume the frequency
to be fixed and equal to the original synchronization frequency $f_{sync}$.
Then, we change $u_{0}$ to explore EPDs out of synchronization, leading
to the results in Fig. \ref{fig: Dispersionu_0}. In this figure,
only wavenumbers with a positive real part are displayed as mentioned
in \cite{wong2018modification}, and a second-order EPD exists around
$u_{0,e}=0.209c$. So the bifurcation is observed in the dispersion
diagram when we select $u_{0}$ larger than $v_{c}$. The bifurcation
of the wavenumber, when $u_{0}$ is varied, is clear evidence of an
EPD.

\begin{figure}[t]
\begin{centering}
\includegraphics[width=2.7in]{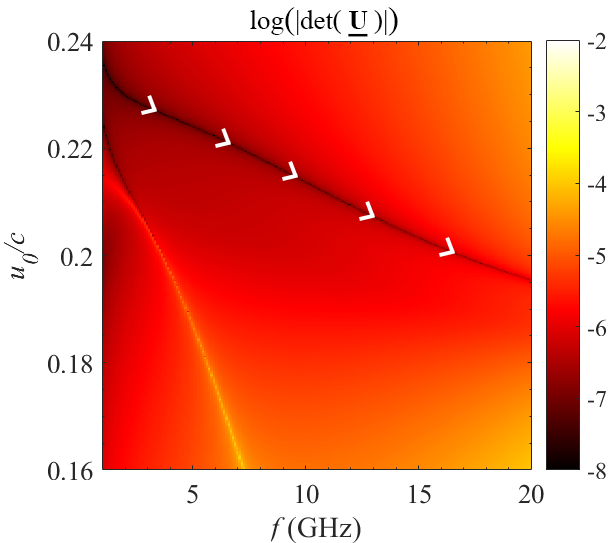}
\par\end{centering}
\centering{}\caption{Determinant of the similarity transformation matrix $\underline{\mathbf{U}}$
varying $f$ and $u_{0}$. The black curve under the white arrows
denotes the location of the exceptional degeneracy.\label{fig: Similarity2D}}
\end{figure}

Based on the results in Figs. \ref{fig: DispersionOmega} and \ref{fig: Dispersionu_0},
we conclude that the TWT-system is very sensitive to variation in
frequency and $u_{0}$ near an EPD. Fig. \ref{fig: Similarity2D}
shows the $\log\left(\left|\det(\mathbf{\underline{U}})\right|\right)$
when frequency and $u_{0}$ are varied. The black curve shows the
lowest values, which means the eigenvectors coalesce at those specific
values of $f$ and $u_{0}$. Thus, the black contour represents EPDs.
EPDs can be utilized to measure e-beam parameters by changing the
frequency in the TWT-system. For a practical scenario, if we have
an e-beam with an unknown $u_{0}$ in the predetermined range, we
can vary the operating frequency to observe EPD and find the corresponding
$u_{0}$.

The eigenvalues at EPDs are extremely sensitive to perturbations of
parameters of the system \cite{zhou2018optical}. Here we establish
that a system's sensitivity to a specific parameter variation is boosted
by the eigenmodes' degeneracy. For instance, let us consider the EPD
in Fig. \ref{fig: DispersionOmega} at $f_{e}=16.82\:\mathrm{GHz}$.
In order to measure the sensitivity of the wavenumber to frequency
variation, the system's relative perturbation parameter is defined
as $\Delta=(f-f_{\mathrm{e}})/f_{\mathrm{e}}$. Consequently, the
perturbed system matrix $\mathbf{\underline{M}}(\Delta)$ has two
degenerate eigenvalues (i.e., the wavenumbers) occurring at the EPD
shift considerably due to a small perturbation in frequency, resulting
in two separate eigenvalues $k_{\mathrm{\mathit{n}}}(\Delta)$, with
$n=1,2$, close to the EPD. These two perturbed eigenvalues are estimated
by using a convergent Puiseux series, where the coefficients are calculated
using the explicit formulas given in \cite{Welters2011Explicit}.
The approximation of $k_{n}(\Delta)$ around a second-order EPD is
given by

\begin{equation}
k_{n}(\Delta)\approx k_{e}+(-1)^{n}\alpha_{1}\sqrt{\Delta}+\alpha_{2}\Delta.\label{eq:Puiseuxm}
\end{equation}

\begin{figure}[t]
\begin{centering}
\includegraphics[width=2.7in]{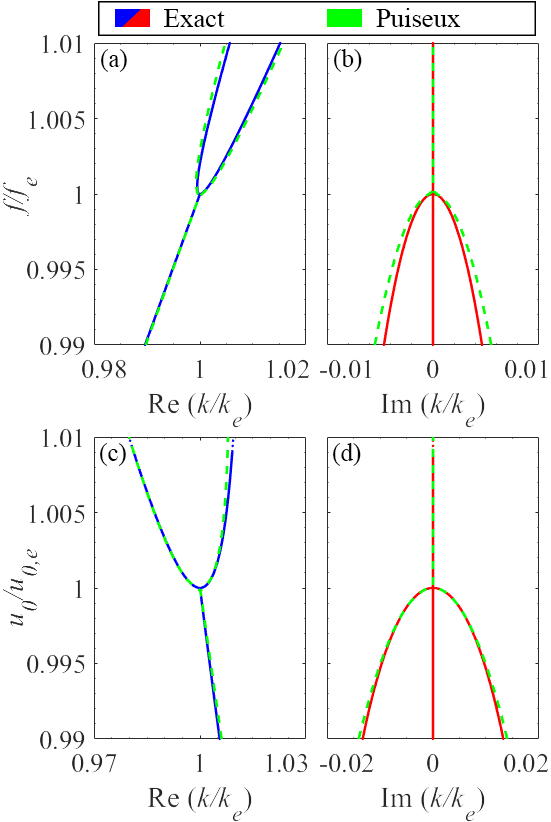}
\par\end{centering}
\centering{}\caption{The Puiseux fractional power expansion up to the second-order approximates
the dispersion diagram variation by frequency and $u_{0}$, further
demonstrating that the bifurcation point is the EPD. (a) and (b) Complex-valued
wavenumbers when frequency is changed, and the required coefficients
are calculated as $\alpha_{1}=99.48\:\mathrm{m^{-1}}$ and $\alpha_{2}=1854.45\:\mathrm{m^{-1}}$.
(c) and (d) Complex-valued wavenumbers when $u_{0}$ is changed, and
the required coefficients are calculated as $\alpha_{1}=175.59\:\mathrm{m^{-1}}$
and $\alpha_{2}=-764.07\:\mathrm{m^{-1}}$.\label{fig: Puiseux}}
\end{figure}

Following \cite{Welters2011Explicit}, $\alpha_{1}$ and $\alpha_{2}$
are calculated by

\begin{equation}
\alpha_{1}=\sqrt{\left(-\frac{\frac{\partial H}{\partial\Delta}(\Delta,k)}{\frac{1}{2!}\frac{\partial^{2}H}{\partial k^{2}}(\Delta,k)}\right)},\label{eq:PuiseuxCoeffm}
\end{equation}

\begin{equation}
\alpha_{2}=-\frac{-(\alpha_{1}^{3}\frac{1}{3!}\frac{\partial^{3}H}{\partial k^{3}}(\Delta,k)+\alpha_{1}\frac{\partial^{2}H}{\partial k\partial\Delta}(\Delta,k))}{2\alpha_{1}(\frac{1}{2!}\frac{\partial^{2}H}{\partial k^{2}}(\Delta,k))},\label{eq: PuiseuxCoeff2m}
\end{equation}
evaluated at the EPD, i.e., at $\Delta=0$ and $k=k_{\mathrm{e}}$,
where $H(\Delta,k)=\mathrm{det}[\underline{\mathbf{M}}(\Delta)-k\underline{\mathbf{I}}]$.
Eq. (\ref{eq:Puiseuxm}) indicates that for a small perturbation $\left|\Delta\right|\ll1$
in frequency, the eigenvalues change dramatically from their original
degenerate value due to the square root dependence. The results in
Figs. \ref{fig: Puiseux}(a) and (b) produce the two branches of the
exact perturbed eigenvalues $k_{n}$ obtained from the eigenvalue
problem when the perturbation $\Delta$ is applied. These figures
explain that such perturbed eigenvalues could be estimated with high
accuracy by using the Puiseux series truncated to its second order.
Next, we analyze the sensitivity to variations in $u_{0}$ by defining
$\Delta=(u_{0}-u_{0,e})/u_{0,e}$ and apply the same procedure to
achieve a Puiseux series coefficients. The calculated results for
perturbation in $u_{0}$ are illustrated in Figs. \ref{fig: Puiseux}(c)
and (d), which demonstrate the bifurcation and high-sensitivity of
the wavenumbers to perturbation near the EPD.

In conclusion, we have investigated the occurrence of EPDs in a system
consisting of a linear e-beam interacting with a guided EM wave. We
have focused on a practical example where the EM wave is guided by
a helix-based SWS, but the same model can be applied to other guiding
geometries. We have considered realistic parameters for the e-beam's
space-charge effect and waveguide's dispersion of phase velocity and
Pierce (interaction) impedance in the developed model. We have discovered
the necessary and required conditions to establish an EPD in TWT-system.
Then, we have discussed how the wavenumbers of the modes participating
in an EPD are extremely sensitive to system perturbations. We have
shown how a bifurcation point well describes such perturbation near
an EPD, also demonstrated by employing the Puiseux fractional power
series expansion. The very high sensitivity to variations can pave
the way to new accurate measurement techniques of e-beam parameters.

\section*{ACKNOWLEDGMENTS}

This material is based upon work supported by the Air Force Office
of Scientific Research award number FA9550-18-1-0355, award number
FA9550-19-1-0103, and award number FA9550-20-1-0409. The authors are
thankful to DS SIMULIA for providing CST Studio Suite that was instrumental
in this study.

\section*{APPENDIX}
\subsection{Electron Beam Model}

There are various approaches to analyzing an electron beam's (e-beam)
interaction with a traveling electromagnetic (EM) wave on a circuit.
The classical small-signal theory by J. R. Pierce is one of the approaches
that still use these days for the modeling and designing of traveling
wave tubes (TWT). Pierce's model includes a qualitative description
of traveling wave interaction that is explained in \cite{pierce1947theory,Pierce1951waves}.
Many of the parameters defined by Pierce are now part of the accepted
vocabulary in traveling wave tube research and industry. For these
reasons, and because the Pierce theory provides considerable physical
insight into TWT mechanics, the basic elements of the Pierce theory
will be described here, together with generalizations that have been
adopted.

In this section, we show the fundamental equations that describe the
e-beam dynamics in both space and time. We follow the linearized equations
that describe the space-charge wave as originally presented by Pierce
\cite{Pierce1951waves}. We assume the e-beam is made of a narrow
cylindrical pencil beam of electrons, which is subject to an axial
(i.e., longitudinal) electric field, assumed constant over the beam's
transverse cross-section; we also consider purely longitudinal electron
motion due to a strong externally applied axial magnetic field which
confines the beam. Because of this, the beam is described by a one-dimensional
function, as will be shown. The beam's total linear charge density
$\rho_{b}^{tot}(z,t)$, and electron velocity $u_{b}^{tot}(z,t)$
are represented as

\begin{equation}
	\rho_{b}^{tot}(z,t)=\rho_{0}+\rho_{b}(z,t),
\end{equation}

\begin{equation}
	u_{b}^{tot}(z,t)=u_{0}+u_{b}(z,t),
\end{equation}
where the subscripts $\lyxmathsym{\textquotedblleft}0\lyxmathsym{\textquotedblright}$
and $\lyxmathsym{\textquotedblleft}b\lyxmathsym{\textquotedblright}$
denote the d.c. (average value) and the a.c. (alternate current, i.e.,
modulation), respectively. In the above equations, $\rho_{0}$ is
negative and $u_{b}^{tot}(z,t)$ is the electron speed in the $z$-direction.
The basic equation that is governing the charges\textquoteright{}
longitudinal motion is

\begin{equation}
	\frac{du_{b}^{tot}(z,t)}{dt}=-\eta e_{z},\label{eq: Charges=002019 Motion}
\end{equation}
where $\eta=e/m=1.758829\times10^{11}C/kg$ is the charge to mass
ratio of an electron, the electron charge is equal to $-e$, and $m$
is the rest mass of the electron. The term $e_{z}$ is the total a.c.
electric field in the $z$-direction, provided by the superposition
of two fields as $e_{z}=e_{w}+e_{p}$, where $e_{w}$ is the $z$-polarization
of the electric field pertaining to the EM mode guided in the waveguide,
and $e_{p}$ is the electric field generated by space-charge, as is
discussed later in this section. Following \cite{Pierce1951waves},
we rewrite the total derivative on the left-hand side of Eq. (\ref{eq: Charges=002019 Motion})
as

\begin{equation}
		\begin{aligned}
	\frac{d(u_{0}+u_{b})}{dt}=\frac{\partial u_{b}}{\partial t}+(u_{0}+u_{b})\frac{\partial u_{b}}{\partial z}+\frac{\partial u_{0}}{\partial t}+(u_{0}+u_{b})\frac{\partial u_{0}}{\partial z}.\\
	\\
		\end{aligned}
\end{equation}

Some terms on this equation vanish because $\partial u_{0}/\partial t=0$,
and $\partial u_{0}/\partial z=0$. Using a small-signal approximation,
we assume that the modulating velocity $u_{b}$ is small with respect
to $u_{0}$; hence the term $u_{b}\partial u_{b}/\partial z$, which
is a product of two a.c. small quantities, is negligible with respect
to the two other terms involving $u_{b}$. Therefore, in our small-signal
theory, we neglect the term $u_{b}\partial u_{b}/\partial z$, as
was originally done by Pierce \cite{Pierce1951waves}. Thus, Eq. (\ref{eq: Charges=002019 Motion})
is rewritten as

\begin{equation}
	\frac{\partial u_{b}(z,t)}{\partial t}+u_{0}\frac{\partial u_{b}(z,t)}{\partial z}=-\eta e_{z}.\label{eq: Charges=002019 Motion2}
\end{equation}

For convenience, we define the equivalent kinetic beam current as

\begin{equation}
	\begin{aligned}
	i_{b}^{tot}(z,t)=u_{b}^{tot}(z,t)\rho_{b}^{tot}(z,t)=u_{0}\rho_{0}+u_{0}\rho_{b}+u_{b}\rho_{0}+u_{b}\rho_{b}\\
	\approx-I_{0}+i_{b}(z,t),\label{eq: i_b,tot}\\
	\\
	\end{aligned}
\end{equation}

Note that we assume a small-signal modulation in the beam speed and
charge density, i.e., we consider a linear model by neglecting the
term $u_{b}\rho_{b}$ in Eq. (\ref{eq: i_b,tot}). Here, the term
$u_{b}\rho_{b}$ is a product of two small a.c. quantities and is
neglected. The a.c. and d.c. portions of the e-beam current are

\begin{equation}
	\left\{ \begin{array}{c}
		i_{b}(z,t)=u_{0}\rho_{b}+u_{b}\rho_{0}\\
		-I_{0}=u_{0}\rho_{0}
	\end{array}\right.\label{eq: Current}
\end{equation}

Moreover, we consider the continuity equation or conservation of charge,

\begin{equation}
	\frac{\partial i_{b}^{tot}(z,t)}{\partial z}=-\frac{\partial\rho_{b}^{tot}(z,t)}{\partial t},
\end{equation}
which is rewritten as follows

\begin{equation}
	\frac{\partial i_{b}(z,t)}{\partial z}-\frac{\partial I_{0}}{\partial z}=-\frac{\partial\rho_{b}(z,t)}{\partial t}-\frac{\partial\rho_{0}}{\partial t}.
\end{equation}

We know that $I_{0}$ and $\rho_{0}$ are d.c. quantities, i.e., their
derivatives are vanishing, so the continuity equation leads finally
to

\begin{equation}
	\frac{\partial i_{b}(z,t)}{\partial z}=-\frac{\partial\rho_{b}(z,t)}{\partial t}.\label{eq: Continuity}
\end{equation}

For a non-relativistic beam, it is convenient to define an equivalent
kinetic beam voltage as

\begin{equation}
	\begin{array}{c}
		v_{b}^{tot}(z,t)=\frac{\left(u_{b}^{tot}(z,t)\right)^{2}}{2\eta}=\frac{u_{0}^{2}+u_{b}^{2}+2u_{0}u_{b}}{2\eta}\approx V_{0}+v_{b}(z,t),\end{array}
\end{equation}
and as explained earlier, based on the small-signal approximation,
we neglect the nonlinear term $u_{b}^{2}$, and separate the a.c.
and d.c. terms as

\begin{equation}
	\left\{ \begin{array}{c}
		v_{b}(z,t)=\frac{u_{0}u_{b}}{\eta}\\
		V_{0}=\frac{u_{0}^{2}}{2\eta}
	\end{array}\right.\label{eq: Voltage}
\end{equation}

By combining Eqs. (\ref{eq: Continuity}), (\ref{eq: Current}), and
(\ref{eq: Voltage}) we find

\begin{equation}
	\frac{\partial i_{b}(z,t)}{\partial z}=\frac{\eta\rho_{0}}{u_{0}^{2}}\frac{\partial v_{b}(z,t)}{\partial t}-\frac{1}{u_{0}}\frac{\partial i_{b}(z,t)}{\partial t}.\label{eq: i_b}
\end{equation}

Moreover, by using Eqs. (\ref{eq: Charges=002019 Motion2}), and (\ref{eq: Voltage})
we write

\begin{equation}
	\frac{\partial v_{b}(z,t)}{\partial z}+\frac{1}{u_{0}}\frac{\partial v_{b}(z,t)}{\partial t}=-e_{z}.\label{eq: Charges=002019 Motion3}
\end{equation}

Eqs. (\ref{eq: i_b}) and (\ref{eq: Charges=002019 Motion3}) are
the two equations governing the e-beam's dynamic based on the model
adopted. In the next step, we elaborate more on the bunching and debunching
effects of the convection beam current in a traveling-wave field.
As was stated previously, the total longitudinal field $e_{z}$ is
represented as the sum of the electric field of the EM mode in the
slow-wave structure (SWS) and the a.c. space-charge field, $e_{z}=e_{w}+e_{p},$where
$e_{w}$ is the $z$-component of the purely vortical field $\mathbf{e}_{w}=\mathrm{curl}\mathbf{b}_{w}$,
where $\mathbf{b}_{w}$ is a magnetic field of the EM mode in the
passive SWS; therefore, $\mathrm{div}\mathbf{e}_{w}=0$ \cite{gewartowski1965principles,Tsimring2007Electron}.
The waveguide EM field $e_{w}$ is provided by \cite{Pierce1951waves}

\begin{equation}
	e_{w}=-a\frac{\partial v}{\partial z},\label{eq:EwPierce}
\end{equation}
where $v$ is the voltage in the equivalent transmission line (TL)
which describes how EM fields propagate in the waveguide, as will
be explained further in the next section. In order to model the interaction
strength between the e-beam and TL, we have generalized the coupling
strength using the coefficient \textit{$a$ }that represents the strength
of interaction between the e-beam and the guided EM mode, as was also
described in \cite{figotin2013multi,tamma2014extension,figotin2020analytic}.
Physically, this coupling strength coefficient describes how strongly
the electric field of a mode in the SWS affects electron motion.

The space-charge field $e_{p}$ is longitudinal, i.e., polarized along
the $z$-direction, and it is generated by electron bunching. It is
determined from the Poisson equation $\nabla\cdot e_{p}=\rho_{v}/\varepsilon_{0}$.
The volumetric charge density $\rho_{v}$ is assumed to be only \textit{$z$}-dependent,
and it is related to the linear charge density by $\rho_{b}=\rho_{v}A$
, where $A$ is the transverse cross-sectional area of the beam. This
leads to

\begin{equation}
	\frac{\partial e_{p}}{\partial z}=\frac{\rho_{b}}{A\varepsilon_{0}}.\label{eq:debunching}
\end{equation}

Differentiating in time on both sides of Eq. (\ref{eq:debunching})
and using Eq. (\ref{eq: Continuity}), the above equation is reduced
to

\begin{equation}
	\frac{\partial^{2}e_{p}}{\partial t\partial z}=-\frac{1}{A\varepsilon_{0}}\frac{\partial i_{b}}{\partial z}.\label{eq:debunching2}
\end{equation}

Now, we rewrite all the equations that will be used to find the hybrid
eigenmodes in the phasor domain assuming implicitly the $\exp(j\omega t)$
time dependence for monochromatic fields. Eq. (\ref{eq: i_b}) is
rewritten in terms of the beam\textquoteright s equivalent voltage
and current phasors as

\begin{equation}
	\frac{\partial I_{b}}{\partial z}=-j\frac{\omega I_{0}}{2V_{0}u_{0}}V_{b}-j\frac{\omega}{u_{0}}I_{b},\label{eq: e-beam - I_b}
\end{equation}
which represents the first of the two main equations that govern the
beam dynamics. The second equation is obtained from Eq. (\ref{eq: Charges=002019 Motion3})
based on the following steps. In the phasor domain, $E_{z}=E_{w}+E_{p}$,
and Eq. (\ref{eq:EwPierce}) is written as $E_{w}(\omega)=-a\left(dV/dz\right)$.
Then, considering the well-known telegrapher's equation in the phasor
domain $dV/dz=-ZI$, where $Z$ is the series per-unit-length TL distributed
impedance, and $I$ is the current in the equivalent TL (see next
section), the longitudinal EM-guided field is found as

\begin{equation}
	E_{w}(\omega)=aZI.\label{eq: E_w}
\end{equation}

By using the phasor form, Eq. (\ref{eq:debunching2}) integrated in
the z-domain is rewritten as

\begin{equation}
	E_{p}(z)=j\frac{1}{A\varepsilon_{0}\omega}I_{b}(z)+\textrm{const.}\label{eq: E_p}
\end{equation}

Finally, using the obtained expression for total longitudinal field
$E_{z}=E_{w}+E_{p}$, Eq. (\ref{eq: Charges=002019 Motion3}) in the
phasor domain yields

\begin{equation}
	\frac{\partial V_{b}}{\partial z}=-aZI-j\frac{\omega}{u_{0}}V_{b}-j\frac{1}{A\varepsilon_{0}\omega}I_{b},\label{eq: V_b}
\end{equation}
which connects the EM mode equivalent current to the e-beam kinetic
voltage and current. This is the second main equation that governs
the beam's dynamics.

So far, we have achieved \textit{two} important first-order linear
differential equations that describe the dynamics of the e-beam kinetic
voltage and current, Eq. (\ref{eq: V_b}) and Eq. (\ref{eq: e-beam - I_b}),
respectively. As we observe in Eq. (\ref{eq: e-beam - I_b}), the
e-beam current is only associated with e-beam parameters, whereas
Eq. (\ref{eq: V_b}) indicates that TL parameters (in this case, $Z$
and $I$) are also required to calculate e-beam voltage. In Eq. (\ref{eq: V_b}),
space-charge fields describe repulsive forces in dense beams of charged
particles. These forces induce oscillations of particles at a plasma
frequency, which, in a moving medium, have the form of a propagating
wave (i.e., the space-charge wave). The plasma frequency is given
by

\begin{equation}
	\omega_{p}=\sqrt{-\frac{\rho_{0}\eta}{A\varepsilon_{0}}}=\sqrt{\frac{I_{0}u_{0}}{2V_{0}A\varepsilon_{0}}}.\label{eq: Omega_p}
\end{equation}

In reality, the beam is enclosed in a metallic structure that affects
the propagation of space-charge waves. Thus, the plasma frequency
of the e-beam, $\omega_{p}$, is effectively decreased, as compared
to its value in the case of an infinite transverse cross section,
to a reduced plasma frequency $\omega_{q}$ \cite{branch1955plasma}.
Therefore, it is important to calculate the reduced plasma frequency
$\omega_{q}$. This is done by accounting for the reduction factor
associated to the plasma frequency $R_{sc}=\omega_{q}/\omega_{p}$
that accounts for the metallic tunnel. In the specific case of a thin
tape helix TWT (HTWT) with a pencil e-beam, Branch and Mihran found
that the helix can be approximated with a perfectly conducting metallic
cylinder of the same internal radius \cite{branch1955plasma}. In
this paper we have used a plasma frequency reduction factor equal
to $R_{sc}=0.12$, which was calculated for the designed SWS using
the software LATTE \cite{antonsen1998traveling,wohlbier2002multifrequency,wohlbier2003latte}.
To better estimate a TWT performance, one simply replaces $\omega_{p}$
with $\omega_{q}$ in the fundamental equations. Therefore, we rewrite
Eq. (\ref{eq: V_b}) as

\begin{equation}
	\frac{dV_{b}}{dz}=-aZI-j\frac{\omega}{u_{0}}V_{b}-j\frac{2V_{0}\omega{}_{q}^{2}}{\omega I_{0}u_{0}}I_{b}.\label{eq: e-beam - V_b}
\end{equation}

The first term on the right-hand side in Eq. (\ref{eq: e-beam - V_b})
shows the role of the electric field of the waveguide EM mode in the
e-beam equations.

\subsection{Electromagnetic Field in the Waveguide Represented by an Equivalent
	Transmission Line and Interaction with the Beam's Charge Wave}

In the TWT-system, the flowing electrons interact with a surrounding
circuit. The convection current in the beam causes current to be induced
in the circuit. This induced current adds to the current already presented
in the circuit, causing the circuit power to increase with distance
as power is extracted from the e-beam. We model the SWS using an equivalent
TL whose equations are

\begin{equation}
	\frac{dV}{dz}=-ZI,
\end{equation}

\begin{equation}
	\frac{dI}{dz}=-YV+i_{s}.
\end{equation}

Here, $V$ indicates the equivalent voltage (related to the electric
field), and $I$ indicates the equivalent current (related to the
magnetic field) in the phasor domain, as explained in \cite{marcuvitz1951representation,felsen1994radiation}.
Furthermore, $Z$ is the distributed series impedance per-unit-length,
and $Y$ is the distributed shunt admittance per-unit-length. In the
above equation, the term $i_{s}$ represents a distributed current
generator \cite{Pierce1951waves,tamma2014extension} that accounts
for the effect of the electron stream flowing in the SWS on the EM
field whose expression is given by $i_{s}=-a\left(dI_{b}/dz\right)$.
We substitute $dI_{b}/dz$ in this latter equation with Eq. (\ref{eq: e-beam - I_b}).
Then, we obtain the set of two fundamental equations for the equivalent
TL

\begin{equation}
	\frac{dV}{dz}=-ZI,\label{eq: TL - V}
\end{equation}

\begin{equation}
	\frac{dI}{dz}=-YV+ja\frac{\omega I_{0}}{2V_{0}u_{0}}V_{b}+ja\frac{\omega}{u_{0}}I_{b}.\label{eq: TL - I}
\end{equation}

In the case of a lossless and non-dispersive waveguide, one has $Z=j\omega L$
and $Y=j\omega C$; however, it is important to note that these equations
are here generalized for realistic lossy and dispersive waveguides
by accounting for the more complex frequency dependence in $Z(\omega)$
and $Y(\omega)$. Indeed, in realistic systems like the one discussed
in this paper, the dispersive waveguide is described by parameters
$Z(\omega)$ and $Y(\omega)$ with nonlinear frequency dependence.
This more involved frequency dispersion can be equivalently accounted
for by defining a dispersive inductance and capacitance per-unit-length
as $Z=j\omega L(\omega)$ and $Y=j\omega C(\omega)$ \cite{wohlbier2002multifrequency,converse2004impulse,setayesh2017pawaic}.

As a final step, we now summarize the system of four equations comprising
the differential equations in Eqs. (\ref{eq: e-beam - I_b}), (\ref{eq: e-beam - V_b}),
(\ref{eq: TL - V}), and (\ref{eq: TL - I}). This system describes
the full dynamics of the linearized (small-signal) model in terms
of the equivalent TL voltage and current, $I(z)$ and $V(z)$, as
well as the charge-wave current and kinetic voltage, $I_{b}(z)$ and
$V_{b}(z)$, respectively. We conveniently define a space-varying
state vector composed of these four EM-field and charge-wave variables
as

\begin{equation}
	\boldsymbol{\Psi}(z)=\left[\begin{array}{c}
		V(z)\\
		\mathbf{\mathit{I}}(z)\\
		V_{b}(z)\\
		I_{b}(z)
	\end{array}\right].
\end{equation}

Without loss of generality, we assume that the TL is homogeneous (i.e.,
\textit{$z$}-invariant), as was originally done by Pierce \cite{Pierce1951waves},
and we write the four fundamental equations in matrix form as

\begin{equation}
	\partial_{z}\boldsymbol{\Psi}(z)=-j\mathbf{\underline{M}}\boldsymbol{\Psi}(z),\label{eq: System}
\end{equation}
where $\mathbf{\underline{M}}$ is a $4\times4$ system matrix \cite{mealy2020exceptional}

\begin{equation}
	\mathbf{\underline{M}}=\begin{array}{c}
		\left[\begin{array}{cccc}
			0 & -jZ & 0 & 0\\
			-jY & 0 & -ag & -a\beta_{0}\\
			0 & -jaZ & \beta_{0} & R_{p}\\
			0 & 0 & g & \beta_{0}
		\end{array}\right]\end{array}.\label{eq: M_Matrix}
\end{equation}

In the above matrix, we have defined the set of parameters as

\begin{equation}
	\beta_{0}=\frac{\omega}{u_{0}},
\end{equation}

\begin{equation}
	g=\frac{1}{2}\frac{I_{0}\beta_{0}}{V_{0}},
\end{equation}

\begin{equation}
	R_{p}=\frac{1}{A\varepsilon_{0}\omega}=\frac{2V_{0}\omega{}_{q}^{2}}{\omega I_{0}u_{0}},\label{eq: R_p}
\end{equation}
where $\beta_{0}$ is beam equivalent propagation constant, and $g$
is a parameter related to the e-beam \cite{mealy2020exceptional}.
In this formulation, we have considered the effect of the bunching
of the convection beam current in a traveling wave field using the
$R_{p}$ term in the above matrix, as was done in \cite{Othman2016Giant,Othman2016Theory}.
This description in terms of a multidimensional first-order differential
equation in Eq. (\ref{eq: System}) is ideal for exploring the occurrence
of an exceptional point of degeneracy (EPD) in the system since an
EPD is a degeneracy associated with two or more coalescing eigenmodes.
In other words, EPDs occur when the system matrix $\mathbf{\underline{M}}$
is similar to a matrix that contains a nontrivial Jordan block. In
general, there are four independent eigenmodes and each eigenmode
is described by an eigenvector $\boldsymbol{\Psi}$.

\subsection{Dispersion Equation}

To obtain the dispersion equation or characteristic equation, we search
for the solution of the form $\boldsymbol{\Psi}(z)=\boldsymbol{\Psi}e^{-jkz}$,
where \textit{$k$} is the complex-valued wavenumber of the hybrid
mode (hybrid because a mode is made of both EM and charge wave components).
The four wavenumbers are obtained by solving

\begin{equation}
		\begin{aligned}
	\det(\mathbf{\underline{M}}-k\mathbf{\underline{I}})=\\
	\det\left[\begin{array}{cccc}
		-k & -jZ & 0 & 0\\
		-jY & -k & -ag & -a\beta_{0}\\
		0 & -jaZ & -k+\beta_{0} & R_{p}\\
		0 & 0 & g & -k+\beta_{0}
	\end{array}\right]=0.
	\end{aligned}
\end{equation}

After some mathematical calculations, the dispersion equation is expressed
as

\begin{equation}
	\begin{array}{c}
		D(\omega,k)=k^{4}-k^{3}(2\beta_{0})+k^{2}(\beta_{0}^{2}-gR_{p}+ZY-ja^{2}Zg)\\
		-k(2ZY\beta_{0})+ZY(\beta_{0}^{2}-gR_{p})=0.\end{array}\label{eq: DispersionSimple}
\end{equation}

Furthermore, the dispersion equation can be rewritten in the convenient
form
\begin{equation}
	(k^{2}+ZY)\left((k-\beta_{0})^{2}-R_{p}g\right)=ja^{2}gZk^{2},
\end{equation}
or we can rewrite it as \cite{tamma2014extension}

\begin{equation}
	(k-\beta_{0})^{2}-\frac{ja^{2}gZk^{2}}{k^{2}+ZY}=gR_{p}.
\end{equation}

The cold circuit phase propagation constant is $\beta_{c}=\sqrt{-ZY},$
and we also used the definition $\beta_{q}=\omega_{q}/u_{0}=\sqrt{R_{p}g}$
which represents the phase constant of the space charge wave traveling
with a phase velocity equal to the average electron velocity and at
an angular frequency equal to $\omega_{q}$. So, the dispersion characteristic
equation is equivalently rewritten as

\begin{equation}
	(k^{2}-\beta_{c}^{2})\left((k-\beta_{0})^{2}-\beta_{q}^{2}\right)=ja^{2}gZk^{2}.
\end{equation}

The right-hand side describes the coupling strength between the two
guiding systems: the wavenumber dispersion in the isolated EM waveguide
(i.e., without e-beam interaction) would be described by $(k^{2}-\beta_{c}^{2})=0$,
and the wavenumber dispersion in the isolated charge wave (i.e., without
interacting with the guided EM wave) would be described by $\left((k-\beta_{0})^{2}-\beta_{q}^{2}\right)=0$.
It is convenient to define the circuit characteristic impedance and
e-beam impedance as

\begin{equation}
	Z_{c}=\frac{Z}{j\beta_{c}}=\sqrt{\frac{Z}{Y}},
\end{equation}

\begin{equation}
	Z_{0}=\frac{V_{0}}{I_{0}}.
\end{equation}

Pierce defined the dimensionless gain parameter $C_{\mathrm{P}}$,
and called it ``gain parameter'' \cite{Pierce1954Traveling},

\begin{equation}
	C_{\mathrm{P}}^{3}=\frac{Z_{c}}{4Z_{0}}.
\end{equation}

Pierce\textquoteright s gain parameter, $C_{\mathrm{P}}$, is a measure
of the intensity of the interaction between the e-beam and SWS. The
characteristic impedance of the equivalent TL $Z_{\mathrm{c}},$ is
also called the interaction impedance or Pierce impedance since it
affects the value of Pierce's gain parameter. Consequently, the above
dispersion equation is rewritten in terms of Pierce's gain parameter
as

\begin{equation}
	(k^{2}-\beta_{c}^{2})\left((k-\beta_{0})^{2}-\beta_{q}^{2}\right)=-2a^{2}C_{\mathrm{P}}^{3}\beta_{c}\beta_{0}k^{2}.\label{eq: Dispersion2}
\end{equation}

It may be convenient to consider a modified Pierce gain parameter
to account for more realistic EM-beam coupling factors due to the
extra coupling strength coefficient we explicitly consider in this
paper, as

\begin{equation}
	C_{\mathrm{P,m}}^{3}=a^{2}C_{\mathrm{P}}^{3}.
\end{equation}

By using this new modified Pierce gain parameter, the dispersion equation
expressed in Eq. (\ref{eq: Dispersion2}) reduces to

\begin{equation}
	(k^{2}-\beta_{c}^{2})\left((k-\beta_{0})^{2}-\beta_{q}^{2}\right)=-2C_{\mathrm{P,m}}^{3}\beta_{c}\beta_{0}k^{2}.
\end{equation}

Note that the term $C_{\mathrm{P,m}}^{3}$ on the right side determines
the coupling strength between the two dispersion equations of the
isolated waveguide and charge-wave guiding systems. If the wavenumber
of the hybrid mode, $k$, in the above equation is solved versus angular
frequency, $\omega$, it is worth recalling that the wavenumber of
the EM wave in the cold SWS, $\beta_{c}$, also depends on frequency
if we consider the waveguide dispersion in our calculations. Furthermore,
the Pierce gain parameter $C_{\mathrm{P,m}}^{3}$ also depends on
frequency when the cold SWS dispersion makes the characteristic impedance
frequency-dependent, aside from the obvious frequency-dependence of
$\beta_{0}$. Furthermore, $R_{sc}$ may also exhibit a slight frequency
variation, though it is assumed constant in this paper based on its
numerical estimation as described in Appendix A.

An alternative description of the hybrid modes is provided in terms
of their phase velocities $v=\omega/k$, rather than their wavenumbers,
as was done in \cite{figotin2020analytic}. Accordingly, the dispersion
equation takes the form of

\begin{equation}
	\frac{(v-u_{0})^{2}}{v^{2}}+\frac{ja^{2}Zgu_{0}^{2}}{v^{2}\beta_{c}^{2}-\omega^{2}}=\frac{\omega_{q}^{2}}{\omega^{2}}.
\end{equation}

\subsection{EPD Condition}

The solutions of our dispersion equations lead to four modal complex-valued
wavenumbers that represent the four hybrid modes in the system. A
second-order EPD occurs when two of these eigenmodes coalesce in their
eigenvalues and eigenvectors, which means that the matrix $\mathbf{\mathbf{\underline{M}}}$
is similar to a matrix that contains a Jordan block of order two \cite{Kato1995Perturbation,mealy2020exceptional}.
In this case, a necessary condition to have second-order EPD is to
have two repeated eigenvalues, which means that the dispersion equation
should have two repeated roots as

\begin{equation}
	D(\omega_{e},k)\propto(k-k_{e})^{2},\label{eq: Second-order EPD}
\end{equation}
where $\omega_{e}$ and $k_{e}$ are the degenerate angular frequency
and wavenumber in EPD condition, respectively. The relation in Eq.
(\ref{eq: Second-order EPD}), which guarantees to have two coalescing
wavenumbers, is satisfied when

\begin{equation}
	D(\omega_{e},k_{e})=0,
\end{equation}

\begin{equation}
	\left.\frac{\partial D(\omega_{e},k)}{\partial k}\right|_{k=k_{e}}=0.
\end{equation}

These two conditions are rewritten, respectively, in the below forms

\begin{equation}
	\begin{array}{c}
		k_{e}^{4}-k_{e}^{3}(2\beta_{0})+k_{e}^{2}(\beta_{0}^{2}-gR_{p,}+Z_{e}Y_{e}-ja^{2}Z_{e}g)\\
		-k_{e}(2Z_{e}Y_{e}\beta_{0})+Z_{e}Y_{e}(\beta_{0}^{2}-gR_{p})=0,\end{array}\label{eq: First Derivation of Dispersion}
\end{equation}

\begin{equation}
	\begin{array}{c}
		4k_{e}^{3}-3k_{e}^{2}(2\beta_{0})+2k_{e}(\beta_{0}^{2}-gR_{p}+Z_{e}Y_{e}-ja^{2}Z_{e}g)\\
		-(2Z_{e}Y_{e}\beta_{0})=0.\end{array}\label{eq: Second Derivation of Dispersion}
\end{equation}

In the above equations, subscript \textquotedbl e\textquotedbl{}
in different parameters indicates the value at the EPD. The TL distributed
series impedance $Z_{e}$, and shunt admittance $Y_{e}$ that provide
the EPD are determined after making some mathematical manipulations
in the two above conditions. First, we use Eq. (\ref{eq: First Derivation of Dispersion})
to get $Y_{e}$ in terms of $Z_{e}$ and other system parameters as

\begin{equation}
	Y_{e}=\frac{-k_{e}^{4}+k_{e}^{3}(2\beta_{0})-k_{e}^{2}(\beta_{0}^{2}-gR_{p}-ja^{2}Z_{e}g)}{Z_{e}\left((k_{e}-\beta_{0})^{2}-gR_{p}\right)},\label{eq: Y_EPD(2)}
\end{equation}
then we substitute this relation into Eq. (\ref{eq: Second Derivation of Dispersion})
and solve it for $Z_{e}$, which is found to be

\begin{equation}
	Z_{e}=\frac{j\left((k_{e}-\beta_{0})^{2}-R_{p}g\right)^{2}}{a^{2}g(-\beta_{0}^{2}+k_{e}\beta_{0}+R_{p}g)}.\label{eq: Z_EPD}
\end{equation}

Finally, we substitute back the impedance value obtained from Eq.
(\ref{eq: Z_EPD}) in the admittance value calculated in Eq. (\ref{eq: Y_EPD(2)})
to find admittance

\begin{equation}
	Y_{e}=\frac{ja^{2}g_{e}k_{e}^{3}(k_{e}-\beta_{0})}{\left((k_{e}-\beta_{0})^{2}-R_{p}g\right)^{2}}.\label{eq: Y_EPD}
\end{equation}

To realize an EPD, the TL series impedance $Z=Z_{e}$ and shunt admittance
$Y=Y_{e}$ need to satisfy Eqs. (\ref{eq: Z_EPD}) and (\ref{eq: Y_EPD}).
Assuming that the EPD conditions in Eqs. (\ref{eq: Z_EPD}) and (\ref{eq: Y_EPD})
are satisfied, then the degenerate wavenumber $k_{e}$ is determined
by the product of Eqs. (\ref{eq: Z_EPD}) and (\ref{eq: Y_EPD})

\begin{equation}
	Z_{e}Y_{e}=\frac{-k_{e}^{3}(k_{e}-\beta_{0})}{(-\beta_{0}^{2}+k_{e}\beta_{0}+R_{p}g)}.
\end{equation}

We know that $\beta_{c,e}^{2}=-Z_{e}Y_{e}$ and $\beta_{q}^{2}=R_{p}g$,
so we calculate $k_{e}$ by solving the equation

\begin{equation}
	\beta_{c,e}^{2}\beta_{q}^{2}=(k_{e}^{3}-\beta_{c,e}^{2}\beta_{0})(k_{e}-\beta_{0}).
\end{equation}

Since we search for solution of the form $\boldsymbol{\Psi}(z)=\boldsymbol{\Psi}_{n}e^{-jk_{n}z}$ , the eigenvectors $\boldsymbol{\Psi}_{n}$ of the system are determined
by solving the eigenvalue problem $\mathbf{\underline{M}}\boldsymbol{\Psi}_{n}=k_{n}\boldsymbol{\Psi}_{n},$or
we can write it as below

\begin{equation}
	(\mathbf{\underline{M}}-k_{n}\mathbf{\underline{I}})\boldsymbol{\Psi}_{n}=0,\label{eq: DispersionDet}
\end{equation}
where $k_{n}$ with $n=1,2,3,4$ are the wavenumbers, and they are
determined from Eq. (\ref{eq: DispersionDet}). By solving Eq. (\ref{eq: DispersionDet}),
the eigenvectors are written in the form of

\begin{equation}
	\boldsymbol{\Psi}_{n}=\left[\begin{array}{c}
		(k_{n}-\beta_{0})^{2}-R_{p}g\\
		j\frac{k_{n}}{Z}\left((k_{n}-\beta_{0})^{2}-R_{p}g\right)\\
		ak_{n}(k_{n}-\beta_{0})\\
		agk_{n}
	\end{array}\right].
\end{equation}

At the second-order EPD investigated in this paper, two of these four
eigenvectors coalesce.

\subsection{Theoretical Gain Calculation}

\begin{figure}[t]
	\begin{centering}
		\includegraphics[width=3.2in]{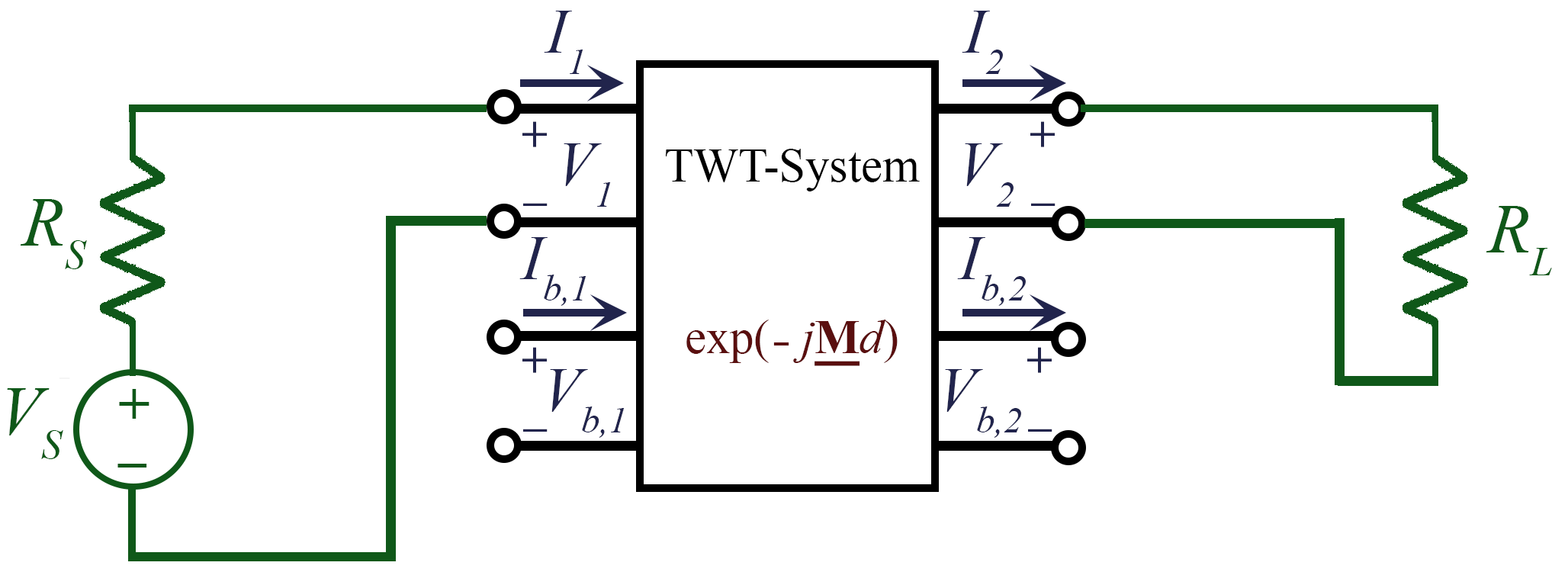}
		\par\end{centering}
	\centering{}\caption{Circuit model that we use for gain calculation.\label{fig: Circuit Model}}
\end{figure}

As we discuss in the main body of the paper, the frequency-dependent
parameters describing EM propagation in the dispersive and lossy waveguide
in the proposed model have a vital role in the accuracy of the calculated
results. In order to test the accuracy of the proposed model, we need
to compare the theoretically calculated results with those numerically
obtained from commercial software. The software LATTE is used to calculate
the gain versus frequency of the helix TWT amplifier. We use our theoretical
method to calculate the power gain versus input signal frequency.
The utilized circuit model is illustrated in Fig. \ref{fig: Circuit Model}.
The TWT-system is modeled by the system matrix $\underline{\mathbf{M}}$,
and we use the input state vector of $\boldsymbol{\Psi}_{1}=[V_{1},I_{1},V_{b,1},I_{b,1}]^{T}$,
calculated at $z=0$, and an output state vector $\boldsymbol{\Psi}_{2}=[V_{2},I_{2},V_{b,2},I_{b,2}]^{T}$
is calculated at $z=d$, i.e., at the end of the SWS. Here $d=Nl$
is the SWS length, where $N$ indicates the number of unit-cells and
$l$ is the SWS period in the $z$-direction. The output state vector
is calculated as

\begin{equation}
	\boldsymbol{\Psi}_{2}=\exp(-j\underline{\mathbf{M}}d)\boldsymbol{\Psi}_{1},
\end{equation}

We considered a helix SWS made of $N=160$ turns, and simulated results
are based on this assumption. In our model, we use the boundary condition
at $z=0$ and $z=d$ provided by the equations

\begin{equation}
	\begin{cases}
		V_{b1}=0\\
		I_{b1}=0\\
		V_{1}+I_{1}R_{S}=V_{S}\\
		V_{2}-I_{2}R_{L}=0
	\end{cases}
\end{equation}

In these equations, the terminations $R_{S}$ (generator resistance)
and $R_{L}$ (load) are assumed to be equal to the frequency-dependent
characteristic impedance of SWS $Z_{c}$ (vary with frequency, to
simulate matching), and $V_{S}$ is the voltage source. Then, we solve
these equations at each frequency and calculate the effective current
and voltage at the output of the TL ($I_{2}$, $V_{2}$). We calculate
the output power $P_{out}=\left|V_{2}\right|^{2}/(2R_{L})$, and the
available input power $P_{avail}=\left|V_{s}\right|^{2}/(8R_{S})$,
to obtain the frequency-dependent gain $G=P_{out}/P_{avail}$, for
the TWT-system. As explained above and in the main body of the paper,
we have introduced the coupling strength coefficient $a$, in our
equations, which describes the strength of beam-EM mode interaction
in the system. The value of $a$ must be optimized in order to obtain
good agreement between the theoretical model and simulation results.
The optimized value for the designed helix TWT is calculated as $a=0.527$.
With this coupling strength coefficient, the theoretical and simulated
gain results are in agreement over the frequency range shown in the
main body of the paper. The agreement between the theoretical and
simulated gain demonstrates the effectiveness of the theoretical model.

\subsection{Comparison to Lagrangian Model}

The Euler-Lagrange equations associated with the Lagrangian are the
following system of second-order differential equations. Without loss
of generality, we rewrote these equations in the case of a single
stream e-beam and a single TL \cite{figotin2013multi,figotin2020analytic}.
All required parameters for this model are summarized in Tables (\ref{tab: Lag e-beam})
and (\ref{tab: Lag TL}), and readers can found more details about
this model in \cite{figotin2020analytic}. The basic equations of
Lagrangian model are represented as

\begin{equation}
	L\partial_{t}^{2}Q-\partial_{z}[C^{-1}(\partial_{z}Q+b\partial_{z}q)]=0,\label{eq: Lagrangian1}
\end{equation}

\begin{equation}
	\frac{1}{\beta}(\partial_{t}+\mathring{v}\partial_{z})^{2}q+\frac{4\pi}{\sigma_{B}}q-b\partial z[C^{-1}(\partial_{z}Q+b\partial_{z}q)]=0.\label{eq: Lagrangian2}
\end{equation}

\begin{table}[tbh]
	\caption{e-beam parameters list in Lagrangian model\label{tab: Lag e-beam}}
	
	\centering{}%
	\renewcommand{\arraystretch}{2}
	\begin{tabular}{|c|c|}
		\hline 
		Name & Value\tabularnewline
		\hline 
		\hline 
		e-beam steady velocity & $\mathring{v}$\tabularnewline
		\hline 
		Number of electron density & $\mathring{n}$\tabularnewline
		\hline 
		Stream intensity & $\beta=\frac{\sigma_{B}}{4\pi}R_{sc}^{2}\omega_{p}^{2}=\frac{R_{sc}^{2}e^{2}}{m}\mathring{N}$\tabularnewline
		\hline 
		Plasma frequency & $\omega_{p}^{2}=\frac{4\pi\mathring{n}_{s}e^{2}}{m}$\tabularnewline
		\hline 
		Plasma frequency reduction factor & $R_{sc}$\tabularnewline
		\hline 
		Beam current & $i$\tabularnewline
		\hline 
		Number of electron per unit of length & $\mathring{N}=\sigma_{B}\mathring{n}$\tabularnewline
		\hline 
		Coupling between e-beam and MTL & $0<b<1$\tabularnewline
		\hline 
		Beam area & $\sigma_{B}$\tabularnewline
		\hline 
	\end{tabular}
\end{table}

\begin{table}[tbh]
	\caption{TL parameters list in Lagrangian model\label{tab: Lag TL}}
	
	\centering{}%
	\renewcommand{\arraystretch}{2}
	\begin{tabular}{|c|c|}
		\hline 
		Name & Value\tabularnewline
		\hline 
		\hline 
		Series inductance per-unit-length & $L$\tabularnewline
		\hline 
		Shunt capacitance per-unit-length & $C$\tabularnewline
		\hline 
		TL characteristic velocity & $w=\frac{1}{\sqrt{LC}}$\tabularnewline
		\hline 
		Coupling coefficient & $b$\tabularnewline
		\hline 
		TL principal coefficient & $\theta=\frac{b^{2}}{C}$\tabularnewline
		\hline 
		TWT principal parameter & $\gamma=\theta\beta=\frac{b^{2}}{C}\frac{\sigma_{B}}{4\pi}R_{sc}^{2}\omega_{p}^{2}$\tabularnewline
		\hline 
	\end{tabular}
\end{table}

In the above equations, $Q(z)$ represents the phasor of the total
amount of a.c. charge flowing through a section at a given $z$ in
the TL, and $L$ and $C$ are the values of inductance and capacitance
associated with the single TL. Also, $q(z)$ represents the amount
of a.c. stream charges modulating the e-beam, at a given section $z$,
in the second-order differential equations. In this section, we wish
to put the equations in the matrix form to solve them. We define the
state vector based on charges in the TL and charges in the e-beam
as

\begin{equation}
	\boldsymbol{\Psi}_{Q}(z)=\left[\begin{array}{c}
		Q(z)\\
		\partial_{z}Q(z)\\
		q(z)\\
		\partial_{z}q(z)
	\end{array}\right].\label{eq: State-Vector Charge}
\end{equation}
Next, we write the Eqs. (\ref{eq: Lagrangian1}) and (\ref{eq: Lagrangian2})
in the matrix form

\begin{equation}
	\partial_{z}\boldsymbol{\Psi}_{Q}(z)=-j\mathbf{\underline{M}}_{QL}\boldsymbol{\Psi}_{Q}(z),\label{eq:largan-system}
\end{equation}
where $\mathbf{\underline{M}}_{QL}$ is a $4\times4$ is system matrix
associated to the Lagrangian formulation and the charges-based state
vector $\boldsymbol{\Psi}_{Q}$, and reads as

\begin{equation}
	\begin{aligned}	
	\mathbf{\underline{M}}_{QL}=\\
	\left[\begin{array}{cccc}
		0 & j & 0 & 0\\
		j(\frac{\beta b^{2}}{\mathring{v}^{2}C}-1)\omega^{2}LC & 0 & -j\frac{b}{\mathring{v}^{2}}(\omega^{2}-\frac{4\pi}{\sigma_{B}}\beta) & -\frac{2\omega}{\mathring{v}}b\\
		0 & 0 & 0 & j\\
		-j\frac{\beta b}{\mathring{v}^{2}}\omega^{2}L & 0 & j\frac{1}{\mathring{v}^{2}}(\omega^{2}-\frac{4\pi}{\sigma_{B}}\beta) & \frac{2\omega}{\mathring{v}}
	\end{array}\right].\label{eq: M_QL}\\
\\
	\end{aligned}
\end{equation}

By defining $Z=j\omega L$ and $Y=j\omega C$, we rewrite Eq. (\ref{eq: M_QL})
as

\begin{equation}
	\begin{aligned}
			\mathbf{\underline{M}}_{QL}=\\
	\left[\begin{array}{cccc}
		0 & j & 0 & 0\\
		j(1-j\omega\frac{\beta b^{2}}{\mathring{v}^{2}Y})ZY & 0 & -j\frac{b}{\mathring{v}^{2}}(\omega^{2}-\frac{4\pi}{\sigma_{B}}\beta) & -\frac{2\omega}{\mathring{v}}b\\
		0 & 0 & 0 & j\\
		-\omega\frac{\beta b}{\mathring{v}^{2}}Z & 0 & j\frac{1}{\mathring{v}^{2}}(\omega^{2}-\frac{4\pi}{\sigma_{B}}\beta) & \frac{2\omega}{\mathring{v}}
	\end{array}\right].\label{eq: First_Charge}\\
\\
	\end{aligned}
\end{equation}

Assuming that our solutions have a $z$-dependence $\boldsymbol{\Psi}(z)=\boldsymbol{\Psi}_{Q}e^{-jkz}$,
the eigenvalue problem reads as,

\begin{equation}
	(\mathbf{\underline{M}}_{QL}-k\mathbf{\underline{I}})\boldsymbol{\Psi}_{Q}=0.
\end{equation}

Based on Eq. (\ref{eq: First_Charge}), the eigenvalue problem is
reduced to

\begin{equation}
	\begin{aligned}
	(\mathbf{\underline{M}}_{QL}-k\mathbf{\underline{I}})\boldsymbol{\Psi}_{Q}=\\
	\left[\begin{array}{cccc}
		-k & j & 0 & 0\\
		j(1-j\omega\frac{\beta b^{2}}{\mathring{v}^{2}Y})ZY & -k & -j\frac{b}{\mathring{v}^{2}}(\omega^{2}-\frac{4\pi}{\sigma_{B}}\beta) & -\frac{2\omega}{\mathring{v}}b\\
		0 & 0 & -k & j\\
		-\omega\frac{\beta b}{\mathring{v}^{2}}Z & 0 & j\frac{1}{\mathring{v}^{2}}(\omega^{2}-\frac{4\pi}{\sigma_{B}}\beta) & -k+\frac{2\omega}{\mathring{v}}
	\end{array}\right]\\
	\boldsymbol{\Psi}_{Q}=0.\\
	\\
	\end{aligned}
\end{equation}

After some simplification, the dispersion equation is expressed as

\begin{equation}
		\begin{aligned}
	\det(\mathbf{\underline{M}}_{QL}-k\mathbf{\underline{I}})=k^{4}-k^{3}(\frac{2\omega}{\mathring{v}})+k^{2}(\frac{1}{\mathring{v}^{2}}(\omega^{2}-\frac{4\pi}{\sigma_{B}}\beta)+ZY\\
	-jb^{2}Z\omega\frac{\beta}{\mathring{v}^{2}})-k(ZY\frac{2\omega}{\mathring{v}})+ZY\frac{1}{\mathring{v}^{2}}(\omega^{2}-\frac{4\pi}{\sigma_{B}}\beta)=0.\label{eq: ChargeDisp1}
		\end{aligned}
\end{equation}

According to the Lagrangian model, the general TWT characteristic
equation for the phase velocity $v=\omega/k$ of the hybrid modes
turns into

\begin{equation}
		\begin{aligned}
	v^{4}\frac{ZY}{\mathring{v}^{2}}(\omega^{2}-\frac{4\pi}{\sigma_{B}}\beta)-v^{3}(ZY\frac{2\omega^{2}}{\mathring{v}})+v^{2}\omega^{2}(\frac{1}{\mathring{v}^{2}}(\omega^{2}-\frac{4\pi}{\sigma_{B}}\beta)\\
	+ZY-jb^{2}Z\omega\frac{\beta}{\mathring{v}^{2}})-v(\frac{2\omega^{4}}{\mathring{v}})+\omega^{4}=0.\label{eq: ChargeDisp1-1}
		\end{aligned}
\end{equation}
After some mathematical manipulation, the characteristic equation
is expressed by \cite[Chapter 25]{figotin2020analytic}

\begin{equation}
	\frac{\gamma}{w^{2}-v^{2}}+\frac{(v-\mathring{v})^{2}}{v^{2}}=\frac{1}{\check{\omega}^{2}},\label{eq: Dispersion-Figotin}
\end{equation}
where $\check{\omega}$ is a dimensionless (normalized) frequency
\cite[Chapter 25]{figotin2020analytic}

\begin{equation}
	\check{\omega}=\frac{\omega}{R_{sc}\omega_{p}}.
\end{equation}

Finally, for convenience we provide the translation table to transform
Lagrangian model parameters used in \cite{figotin2020analytic} to
the Pierce model parameters used in this paper. The list of transformations
is summarized in Table (\ref{tab: Translation}).

\begin{table}[tbh]
	\caption{Translation from Lagrangian model parameters to the Pierce model parameters\label{tab: Translation}}
	
	\centering{}%
	\renewcommand{\arraystretch}{2}
	\begin{tabular}{|c|c|}
		\hline 
		Lagrangian model & Pierce model\tabularnewline
		\hline 
		\hline 
		$\mathring{v}$ & $\frac{\omega}{\beta_{0}}=u_{0}$\tabularnewline
		\hline 
		$\sigma_{B}$ & $A$\tabularnewline
		\hline 
		$\beta$ & $\frac{g\omega}{\beta_{0}^{2}}=\frac{gu_{0}}{\beta_{0}}$\tabularnewline
		\hline 
		$w^{2}$ & $-\frac{\omega^{2}}{ZY}=\frac{\omega^{2}}{\beta_{c}^{2}}$\tabularnewline
		\hline 
		$\gamma$ & $\frac{a^{2}}{Y}\frac{jg\omega^{2}}{\beta_{0}^{2}}=\frac{a^{2}}{Y}jgu_{0}^{2}$\tabularnewline
		\hline 
	\end{tabular}
\end{table}

In the frequency-dependent SWS model that we have introduced in this
paper, we consider two frequency-dependent parameters, i.e., the cold
circuit EM phase velocity $v_{c}$, and the equivalent TL characteristic
impedance $Z_{c}$. The same procedure can be used for the demonstrated
Lagrangian model in \cite{figotin2020analytic}. In the Lagrangian
model, the TL principal coefficient, $\theta(\omega)$, and TL characteristic
velocity, $w(\omega)$, are the two frequency-dependent parameters
in the Lagrangian equations. Equivalently, in the displayed characteristic
equation in Eq. (\ref{eq: Dispersion-Figotin}), the TWT principal
parameter, $\gamma(\omega)$, and TL characteristic velocity, $w(\omega)$,
are the two frequency-dependent parameters in the Lagrangian model.

\subsection{Space-Charge Effect}

Space-charge fields represent repulsive forces in dense beams of charged
particles. These forces induce oscillations of particles at a plasma
frequency, which, in a moving medium, has the form of a propagating
wave (i.e., the space-charge wave). In this paper, we have provided
TWT-system equations which account for space-charge effects. This
effect can be modeled based on calculations provided in the first
section of Appendix for the Lagrangian model of
the TWT-system in \cite{figotin2013multi,figotin2020analytic}. In
the first step, we start with the extended equations represented in
\ref{eq: System},

\begin{equation}
	\begin{cases}
		\partial_{z}V=-ZI\\
		\partial_{z}I=-YV+ja\frac{\omega I_{0}}{2V_{0}u_{0}}V_{b}+ja\frac{\omega}{u_{0}}I_{b}\\
		\partial_{z}V_{b}=-aZI-j\frac{\omega}{u_{0}}V_{b}-j\frac{1}{\omega A\varepsilon_{0}}I_{b}\\
		\partial_{z}I_{b}=-j\frac{\omega I_{0}}{2V_{0}u_{0}}V_{b}-j\frac{\omega}{u_{0}}I_{b}
	\end{cases}
\end{equation}

Then, we transform the four first-order differential equations into
two second-order differential equations by removing voltages, $V$
and $V_{b}$, leading to

\begin{equation}
	\begin{cases}
		\partial_{z}^{2}I-ZYI+a\partial_{z}^{2}I_{b}=0\\
		jag\partial_{z}^{2}I-Y\partial_{z}^{2}I_{b}-j2Y\beta_{0}\partial_{z}I_{b}+Y\beta_{0}^{2}I_{b}-gYR_{p}I_{b}\\
		+ja^{2}g\partial_{z}^{2}I_{b}=0
	\end{cases}
\end{equation}

In the next step, we use the below substitutions for converting currents
to charges

\begin{equation}
	\left\{ \begin{array}{c}
		I=j\omega Q\\
		I_{b}=j\omega q
	\end{array}\right.\label{eq: I_to_Q}
\end{equation}

After some mathematical manipulation, we obtain these equations

\begin{equation}
	\begin{cases}
		L\partial_{t}^{2}Q-\partial_{z}[C^{-1}(\partial_{z}Q+a\partial_{z}q)]=0\\
		\frac{\beta_{0}^{2}}{g\omega}(\partial_{t}+\frac{\omega}{\beta_{0}}\partial_{z})^{2}q+\omega R_{p}q-a\partial_{z}[C^{-1}((\partial_{z}Q+a\partial_{z}q))]=0\\
	\end{cases}\label{eq: Pierce_LagrangianForm}
\end{equation}

This set of equations are equivalent to Euler-Lagrange equations,
which are presented in Eqs. (\ref{eq: Lagrangian1}), and (\ref{eq: Lagrangian2}).
The term $\omega R_{p}q$ is responsible for the space-charge effect.
It can also be written as

\begin{equation}
	\omega R_{p}q=\frac{1}{A\varepsilon_{0}}q.\label{eq: OmegaR_pq}
\end{equation}

On the other hand, in the presented Euler-Lagrange equations, the
term $\frac{4\pi}{\sigma_{B}}q$, accounts for the debunching effect
(See Eq. (\ref{eq: Lagrangian2})). Since the Gaussian system of units
is utilized in the Euler-Lagrange equations, we need to transform
parameters to the International System of Units (SI). After performing
the mentioned transformation, we obtain the same definition as presented
in Eq. \ref{eq: OmegaR_pq}.

\subsection{An Equivalent Alternative Formulation Based on Charge}

As explained in the previous section, we defined matrix equations
for the TWT-system as expressed in Eq. (\ref{eq: M_Matrix}). In the
next step, we start with the substitutions presented in Eq. \ref{eq: I_to_Q}
to convert currents to charges. This leads to modified set of TWT-system
equations as

\begin{equation}
		\begin{aligned}
	\partial_{z}\left[\begin{array}{c}
		\mathbf{\mathrm{\mathit{V}}}\\
		\mathbf{\mathit{j\omega Q}}\\
		V_{b}\\
		j\omega q
	\end{array}\right]=\left[\begin{array}{cccc}
		0 & -Z & 0 & 0\\
		-Y & 0 & jag & ja\beta_{0}\\
		0 & -aZ & -j\beta_{0} & -jR_{p}\\
		0 & 0 & -jg & -j\beta_{0}
	\end{array}\right]\left[\begin{array}{c}
		\mathbf{\mathrm{\mathit{V}}}\\
		\mathbf{\mathit{j\omega Q}}\\
		V_{b}\\
		j\omega q
	\end{array}\right].\\
\\
	\end{aligned}
\end{equation}

The matrix is equivalent to the four equations

\begin{equation}
	\left\{ \begin{array}{c}
		\partial_{z}V=-jZ\omega Q\\
		j\omega\partial_{z}Q=-YV+jagV_{b}-a\beta_{0}\omega q\\
		\partial_{z}V_{b}=-jaZ\omega Q-j\beta_{0}V_{b}+R_{p}\omega q\\
		j\omega\partial_{z}q=-jgV_{b}+\beta_{0}\omega q
	\end{array}\right.
\end{equation}

By combining the equations and performing some mathematical simplification,
we remove voltages ($V$ and $V_{b}$) from equations and decrease
four first-order differential equations into two second-order differential
equations based on charge:

\begin{equation}
	\left\{ \begin{array}{c}
		\partial_{z}^{2}Q=YZQ-ja^{2}gZQ+ja\beta_{0}\partial_{z}q-a\beta_{0}^{2}q+agR_{p}q\\
		+ja\beta_{0}\partial_{z}q\\
		\partial_{z}^{2}q=jagZQ-j\beta_{0}\partial_{z}q+\beta_{0}^{2}q-gR_{p}q-j\beta_{0}\partial_{z}q
	\end{array}\right.\label{eq: Second-order Q}
\end{equation}

In order to analyze the characteristics of the system like wavenumbers,
we rewrite equations in the matrix form. So, we use a state vector
based on charge, as was expressed before in Eq. (\ref{eq: State-Vector Charge}),
and rewrite Eq. (\ref{eq: Second-order Q}) as

\begin{equation}
	\partial_{z}\boldsymbol{\Psi}_{Q}(z)=-j\mathbf{\underline{M}}_{Q}\boldsymbol{\Psi}_{Q}(z),
\end{equation}

\begin{equation}
		\begin{aligned}
	\mathbf{\underline{M}}_{Q}=\left[\begin{array}{cccc}
		0 & j & 0 & 0\\
		jYZ+a^{2}gZ & 0 & -ja\beta_{0}^{2}+jagR_{p} & -2a\beta_{0}\\
		0 & 0 & 0 & j\\
		-agZ & 0 & j\beta_{0}^{2}-jgR_{p} & 2\beta_{0}
	\end{array}\right].\\
\\
	\end{aligned}
\end{equation}

By using the same approach described before, the characteristic equation
is calculated from

\begin{equation}
		\begin{aligned}
	\det(\mathbf{\underline{M}}_{Q}-k\mathbf{\underline{I}})=\\
	\det\left[\begin{array}{cccc}
		-k & j & 0 & 0\\
		jYZ+a^{2}gZ & -k & -ja\beta_{0}^{2}+jagR_{p} & -2s\beta_{0}\\
		0 & 0 & -k & j\\
		-agZ & 0 & j\beta_{0}^{2}-jgR_{p} & -k+2\beta_{0}
	\end{array}\right]\\
=0,\\
\\
	\end{aligned}
\end{equation}

resulting in the following dispersion equation

\begin{equation}
		\begin{aligned}
	D(\omega,k)=k^{4}-k^{3}(2\beta_{0})+k^{2}(\beta_{0}^{2}-gR_{p}+ZY-a^{2}jZg)\\
	-k(2ZY\beta_{0})+ZY(\beta_{0}^{2}-gR_{p})=0.\\
	\\
		\end{aligned}
\end{equation}

\bibliographystyle{IEEEtran}
\bibliography{Ref}

\end{document}